\documentstyle[aps]{revtex}
\begin{document}
\newcommand{\beq}{\begin{equation}}
\newcommand{\eeq}{\end{equation}}
\newcommand{\beqn}{\begin{eqnarray}}
\newcommand{\eeqn}{\end{eqnarray}}
\newcommand{\bmath}{\begin{mathletters}}
\newcommand{\emath}{\end{mathletters}}
\twocolumn[\hsize\textwidth\columnwidth\hsize\csname @twocolumnfalse\endcsname 
\title{ Quasiparticle undressing in a dynamic Hubbard model: exact diagonalization study}
\author{J. E. Hirsch }
\address{Department of Physics, University of California, San Diego\\
La Jolla, CA 92093-0319}

\date{\today} 
\maketitle 
\begin{abstract} 
Dynamic Hubbard models have been proposed as  extensions of the conventional
Hubbard model to describe the orbital relaxation that occurs upon double
occupancy of an atomic orbital. These models give rise to pairing of holes and
superconductivity in certain parameter ranges. Here we explore   the changes in
carrier effective mass and quasiparticle weight
 and in one- and two-particle spectral functions that occur
in a dynamic Hubbard model upon pairing, by exact diagonalization of small systems. 
It is found that pairing is associated with lowering of
effective mass and increase of quasiparticle weight, manifested in
 transfer of spectral weight from high to low frequencies in one- and two-particle
spectral functions. This 'undressing' phenomenology resembles observations in
transport, photoemission and optical experiments in high $T_c$ cuprates. This
behavior is contrasted with that of a conventional electron-hole symmetric 
Holstein-like model  with
attractive on-site interaction, where pairing is associated with
'dressing' instead of 'undressing'.
\end{abstract}
\pacs{}

\vskip2pc]
 
\section{Introduction}
The conventional understanding of superconductivity starts from a normal
state composed of weakly interacting 'dressed' Landau quasiparticles\cite{schrieffer}. 
When the system
goes superconducting these quasiparticles become correlated in Cooper pairs.
As a function of increasing coupling strength, the Cooper pairs undergo
a crossover from weakly bound with a long coherence length to strongly
bound with a short coherence length. Because of the increased correlation
of the quasiparticles in the Cooper pair their effective mass increases
and their quasiparticle weight decreases compared to the normal state,
the more so the stronger the coupling. In other words, the quasiparticle
'dressing' is larger in the superconducting than in the normal state.

Instead, the theory of hole superconductivity\cite{hole1,hole2,undressing} proposes a new 
paradigm to describe superconductivity which is exactly opposite to what
is described above. It also starts from a Fermi liquid of weakly
interacting dressed quasiparticles, and Cooper pairs are also formed as the system
goes superconducting. However, here the quasiparticles  'undress' and resemble more 
free particles when they are bound in the Cooper pair than when they are unbound in the normal
state, with the effect being largest precisely in the strong coupling short
coherence length regime.

This paradoxical scenario is described by a new class of model Hamiltonians
recently introduced to describe correlated electrons, 
'dynamic Hubbard models'.\cite{hole1,dynh1,dynh2}
It is argued that these models capture an essential aspect of the physics of
correlated electrons in atoms, molecules and solids, which is left out in other
models like the conventional  Hubbard model. Whether these new models
describe superconductivity in any or all real materials remains to be established.
The fundamental feature distinguishing these from conventional
models is that they describe 'undressing' instead of 'dressing' when carriers
pair. The undressing should be most apparent for superconductors
with high critical temperature. In support of these models over conventional
models we remark that many aspects of the phenomenology of high $T_c$
cuprate superconductors, to be reviewed later, indicate that quasiparticles undress when they 
pair\cite{ding,ando,johnson,basov,santander,marel}.
Of course there could be other unconventional models describing similar
physics.

In a recent paper we have begun a numerical study of a particular realization
of a dynamic Hubbard model with auxiliary spin degrees of freedom.\cite{qmc}
We studied the pair binding energy and lowering of kinetic energy that occurs when 
carriers pair as function of the parameters in the model
by exact diagonalization of small clusters and quantum Monte Carlo simulations, and obtained
the approximate phase diagram of the model in one dimension. The purpose of this paper 
is to learn more about the properties of this model by studying 
frequency-dependent spectral functions. The reason for doing so is that
the properties of this simple model are likely to be generic for the entire  class of dynamic Hubbard
models and, as argued elsewhere\cite{dynh2}, representative of the properties of correlated electrons
in solids.  We calculate the frequency
dependent conductivity and the single particle spectral function  in this model by
exact diagonalization of small clusters, as well as the quasiparticle weight and
effective mass, and examine their behavior as
function of parameters. The results support the general scenario of
'undressing' in these models indicated by more approximate treatments and
qualitative arguments. We contrast the behavior of this model with that
of a conventional (electron-hole symmetric) model where 'dressing' rather than 'undressing' occurs upon
pairing. Finally, we discuss the connection of our results to experimental
observations in spectroscopic experiments in high $T_c$ materials.

\section{Models} 
\subsection{Dynamic Hubbard model}
The dynamic Hubbard model of interest here is defined by the Hamiltonian\cite{dynh2,qmc}
\beq
H=\sum_i H_i -t\sum_{i,\sigma} [c_{i\sigma}^\dagger c_{i+1,\sigma} + h.c.]
\eeq
where the local Hamiltonian $H_i$ in $electron$ representation is
\bmath\beq
H_i=\omega_0 \sigma_x^i + g \omega_0 \sigma_z^i
+[U-2g\omega_0\sigma_z^i]n_{i\uparrow}n_{i\downarrow}
\eeq
and in hole representation
\beqn
H_i&=&\omega_0 \sigma_x^i+ g \omega_0 [2(n_{i\uparrow}+n_{i\downarrow})-1]\sigma_z^i
\nonumber \\
& &+[U-2g\omega_0\sigma_z^i]n_{i\uparrow}n_{i\downarrow}
\eeqn
\emath
and $\sigma_x^i, \sigma_z^i$   are Pauli matrices associated with an auxiliary 
spin-1/2 degree of freedom at each site.
In what follows the Hamiltonian in hole representation will be used.
$U$ is the effective on-site interaction. Briefly, the auxiliary spin is introduced
to allow for the fact that two electrons on a site can be in more than one state:
depending on the orientation of the auxiliary spin, they will experience more
or less Coulomb repulsion and in turn pay less or more in kinetic and single-electron
potential energy, described by the energy of the auxiliary spin. A more detailed
 justification of the site Hamiltonian Eq. (2) 
to describe the physics of real atoms is discussed in ref. \cite{qmc}.

In the antiadiabatic limit $\omega_0\rightarrow \infty$ the effective low energy
Hamiltonian for low hole concentration is\cite{qmc}
\bmath
\beqn
H_{eff}&=&-\sum_{i,\sigma}[t_2+\Delta t(n_{i,-\sigma}+n_{i+1,-\sigma})]
(c_{i\sigma}^\dagger c_{i+1\sigma}+h.c.) \nonumber \\
& &+U\sum_i n_{i \uparrow}n_{i \downarrow}
\eeqn
\beq
t_2=S^2t
\eeq
\beq
\Delta t=tS(1-S)
\eeq
\beq
S=\frac{1}{\sqrt{1+g^2}} .
\eeq
\emath
which describes ground-state to ground-state transitions of the spin degree of
freedom at each site when the holes hop. This Hamiltonian gives rise to pairing of 
two holes in a full band if the condition 
\beq
\frac{U}{4t}\leq \frac{g^2}{1+g^2}
\eeq
is satisfied.
As shown in Ref\cite{qmc}, the condition for pairing for finite $\omega_0$ is
considerably less stringent than Eq. (4).

\subsection{Electron-hole symmetric Holstein-like model}
Conventional electron-boson models  involve
coupling of a boson (e.g. phonon) degree of freedom to the electronic charge density,
and are electron-hole symmetric. We will contrast the behavior of the
dynamic Hubbard model  with that of a model with
site Hamiltonian
\beq
H_i=\omega_0 \sigma_x^i+g\omega_0 \sigma_z^i 
[n_{i\uparrow}+n_{i\downarrow}-1]+U_0n_{i\uparrow}n_{i\downarrow}
\eeq
as a generic 'conventional' model. Some properties of this Hamiltonian
were discussed in ref. \cite{qmc}. The effective on-site interaction is
\beq
U=U_0-2\omega_0 (\sqrt{1+g^2}-1) .
\eeq
The low energy effective Hamiltonian in the antiadiabatic limit
with $\omega_0\rightarrow\infty$, $g$ fixed and $U$ fixed (i.e. $U_0\rightarrow \infty$
also) is
\bmath
\beq
H_{eff}=-t_{eff}\sum_{i,\sigma}(c_{i\sigma}^\dagger c_{i+1\sigma}+h.c.) 
+U\sum_i n_{i \uparrow}n_{i \downarrow}
\eeq
with 
\beq
t_{eff}=tS^2
\eeq
\beq
S=<0|1>=<1|2>=\sqrt{\frac{1}{2}(1+ \frac{1}{\sqrt{1+g^2}})}
\eeq
\emath
and it gives rise to pairing only if $U<0$ (attractive Hubbard model). For finite
$\omega_0$ we find that the condition for pairing is even more
stringent than $U<0$. In contrast to the dynamic Hubbard model, here 
$S$ does not become small as $g\rightarrow \infty$.

\section{Spectral functions}
\subsection{Optical conductivity}

We will compute the optical conductivity in these models at zero
temperature, given by
\beq
\sigma_1(\omega)=\pi\sum_m\frac{|<0|J|m>|^2}{E_m-E_0}\delta(\omega-(E_m-E_0))
\eeq
with the current operator given by
\beq
J=it\sum_i[c_{i+1\sigma}^\dagger c_{i\sigma}-c_{i\sigma}^\dagger c_{i+1\sigma}]
\eeq
It is easily shown following the steps in Maldague's derivation\cite{maldague}
 that the sum rule
\beq
\int_0^\infty d\omega \sigma_1(\omega)=\frac{\pi}{2}<0|-T|0>
\eeq
holds for these models, with $T$ the bare kinetic energy operator
\beq
T= -t\sum_{i,\sigma} [c_{i\sigma}^\dagger c_{i+1,\sigma} + h.c.]   .
\eeq

When the frequency $\omega_0$ is not very small there is a natural separation
of energy scales in the Hamiltonians. A low-lying submanifold of states in
the Hilbert space corresponds to states where each site spin is at its site
ground state, which is different depending on the electronic occupation\cite{qmc}.
The effective Hamiltonian Eq. (3) or Eq. (7) describes the coupling between those 
low-lying states,
where the electrons hop from site to site and the spins make ground-state
to ground-state (diagonal) transitions. This part of the Hilbert space
corresponds to the quasiparticle band, and the optical absorption involving
transitions between those states  is the low frequency 'intraband' part
of $\sigma_1$, corresponding to the Drude part of the optical 
conductivity.
We can then decompose the integral of the optical conductivity as
\beq
\int_0^\infty d\omega \sigma_1(\omega)=\int_0^{\omega_m}d\omega \sigma_1(\omega)
+\int_{\omega_m}^\infty d\omega \sigma_1(\omega)\equiv A_l+A_h
\eeq
where $A_l$ is the intra-band absorption and $\omega_m$ is a frequency
cutoff that restricts optical transitions to the subset  of intraband states.
On the other hand, an optical sum rule also holds for the effective Hamiltonians
\beq
\int_0^\infty d\omega \sigma_1^{eff}(\omega)=\frac{\pi}{2}<0|-T_{eff}|0>
\eeq
where 
\beq
T_{eff}=-\sum_{i,\sigma}[t_2+\Delta t (n_{i,-\sigma}+n_{i+1,-\sigma})]
(c_{i\sigma}^\dagger c_{i+1\sigma}+h.c.)]
\eeq
for the site Hamiltonian Eq. (2) and 
\beq
T_{eff}=-t_{eff}\sum_{i,\sigma}(c_{i\sigma}^\dagger c_{i+1\sigma}+h.c.) 
\eeq
for the site Hamiltonian Eq. (5), 
and where $\sigma_1^{eff}$ is computed from Eq. (8) with the eigenstates
and eigenvalues of the effective Hamiltonians. Because the 
low-lying spectrum and eigenstates of the full Hamiltonians coincide
with those of the effective Hamiltonians  we can write Eq. (13) as a
'partial' conductivity sum rule for the low-lying eigenstates and 
eigenvalues of the full Hamiltonian
\beq
\int_0^{\omega_m} d\omega \sigma_1(\omega)=A_l=\frac{\pi}{2}<0|-T_{eff}|0>
\eeq
The ground state $|0>$ on the right side of Eq. (16) is the ground state of
the effective Hamiltonian Eq. (3) or Eq. (7).

For the effective Hamiltonian of the dynamic Hubbard model Eq. (3) it can
be seen within BCS theory\cite{apparent}, as well as from the exact solution in the dilute
limit\cite{london} that when pairing occurs the right-hand side of Eq. (16)
increases. This extra spectral weight signals a lowering of kinetic energy
and effective mass reduction in the quasiparticle band. In a real physical
system the total integrated optical spectral weight is conserved, so that any
extra spectral weight at low frequency has to come at the expense of spectral
weight in another frequency range. In tight binding models however the
total optical spectral weight is not conserved, because the current and kinetic
energy operators do not describe transitions to states in other bands. As a 
consequence, the total integral Eq. (12) does not remain constant but also
increases upon pairing in our model. Nevertheless, we will see in the 
numerical results that the dominant effect of pairing in the dynamic Hubbard
model is a large increase
in the low frequency 'intra-band' spectal weight $A_l$ and an overall
shift in optical spectral weight from higher to lower frequencies.

In contrast, we will see that
in the electron-hole symmetric model, the behavior is exactly
opposite: in the paired state the effective mass of the carriers increases
hence optical spectral weight is transfered from the quasiparticle band
to higher frequencies.

\subsection{One-particle spectral function}

The changes in the optical conductivity upon pairing are intimately
related to changes in the one-particle spectral functions. We consider
here the spectral function for hole destruction in a system of $n+1$ holes,
defined as\cite{dynh2}
\bmath
\beq
A_{n+1,n}(\omega)=\sum_l |<l_n|c_{k\sigma}|0_{n+1}>|^2\delta(\omega-(E_l^n-E_0^n))
\eeq
as well as the spectral function for hole creation in a system of $n$ holes:
\beq
A_{n,n+1}(\omega)=\sum_l |<l_{n+1}|c^\dagger_{k\sigma}|0_n>|^2
\delta(\omega-(E_l^{n+1}-E_0^n))
\eeq
\emath
Here, $|l_n>$ denotes the $l-$th excited state of the system with $n$ holes.
In the exact diagonalization calculation we will study clusters with one hole
and two holes of opposite spin. The relevant momentum is then $k=0$. 

Consider these functions for a single site in the dynamic Hubbard model. They are given by
\bmath
\beq
A_{21}(\omega)=A_{12}(\omega)=\delta(\omega)
\eeq
\beq
A_{10}(\omega)=A_{01}(\omega)=
S^2\delta(\omega)+(1-S^2)\delta(\omega-2\omega_0\sqrt{1+g^2})
\eeq
\emath
with $S$ given by Eq. (3d).
The $\omega=0$ part corresponds to the quasiparticle contribution, and its
coefficient is the quasiparticle weight $z$. It can be seen that for the
single site, $z=1$ for the spectral function involving one and two holes
 and $z<1$ for the spectral function involving zero and one hole. In an
extended system, when $z<1$ there will be an 'incoherent' contribution
to the spectral function represented in the site by the second term in 
Eq. (18b). Hence we expect in an extended system that in the unpaired
state the quasiparticle weight will be small, and that when pairing
occurs spectral weight in the single particle spectral function will be 
transfered from the high frequency incoherent
part to the $\omega=0$ quasiparticle peak. Correspondingly,
optical transitions involving transitions between singly and doubly
occupied sites will have larger low energy spectral weight than those
involving empty and singly occupied sites, and optical spectral
weight should be transfered from high frequencies to low frequencies
when pairing occurs.

In contrast, in the electron-hole symmetric model the  spectral
functions for creation of a hole in a singly occupied site and destruction
of a hole in a singly occupied site are both given by
\bmath
\beq
A_{12}(\omega)=A_{10}(\omega)=S^2\delta(\omega)+(1-S^2)\delta(\omega-2\omega_0\sqrt{1+g^2})
\eeq
and the spectral functions for destruction of a hole in a doubly occupied site
and creation of a hole in an empty site are given by
\beq
A_{21}(\omega)=A_{01}(\omega)=S^2\delta(\omega)+(1-S^2)\delta(\omega-2\omega_0)
\eeq
\emath
where $S$ is given by Eq. (7c). It can be seen that the quasiparticle weight is 
the same for all these spectral functions ($z=S^2$), whether the site is initially empty,
singly or doubly occupied. Hence we cannot extract any conclusions about
changes in the quasiparticle weight upon pairing from 'single site' physics
in this model.
In an extended system however because of the more correlated nature of the
wavefunction in the paired state we will see that pairing is associated with
decrease of the quasiparticle weight, i.e. increased 'dressing', in contrast to 
the behavior in the dynamic Hubbard model. Similarly one would expect transfer
of optical spectral weight from low frequencies to high frequencies upon
pairing in this model, in contrast to the behavior in the dynamic Hubbard model.

\section{Numerical results}

We diagonalize exactly the Hamiltonian Eq. (1) with site Hamiltonians Eq. (2b) (dynamic
Hubbard model) and Eq. (6) (electron-hole symmetric model) for four sites in the 
subspaces with zero-, one- and two-hole occupation. We choose units so that
$t=1$. For a finite chain, the optical sum rule Eq. (10) holds if free ends
boundary conditions are use, but not if periodic boundary conditions are used; in the
latter case, an extra term proportional to a zero-frequency $\delta$-function is needed
to satisfy the sum rule\cite{hanke}. We will use free ends boundary conditions to calculate
the optical conductivity, so that the total optical spectral weight is obtained
in the numerical calculation. The lowest frequency peak then occurs at a finite
frequency $\omega_D$ ('Drude 'precursor')\cite{hanke} that goes to zero 
as the size of the system increases.

\subsection{Dynamic Hubbard model}

To obtain a clear separation of energy scales that illustrates clearly the
physics of the model we choose a strong coupling case, with $g=3$.
The single hole effective hopping Eq. (3b) in the antiadiabatic limit is
then $t_2=0.1$, and $\Delta t=0.216$.
 The site energies for $n$ holes (excluding the $U$ term) are given by
\beq
\epsilon(n)=-\epsilon(\bar{n})=-\omega_0\sqrt{1+g^2}
\eeq
with $\epsilon(n), \epsilon(\bar{n})$ the energies of the site ground state and
excited state. Hence the difference $\epsilon(n)-\epsilon(\bar{n})=6.3\omega_0$ is
much larger than the effective bandwidth for the holes except for very small
$\omega_0$.  Even so, for the case of 2 holes the
effective Hamiltonian Eq. (3) is not an accurate representation for finite
frequencies because of the contributions from 'vertical transitions' as discussed
in ref. \cite{qmc}; the effective interaction
\beq
U_{eff}=E(2)+E(0)-2E(1)
\eeq
($E(n)=$energy with $n$ holes) is considerably more attractive for finite
frequency than in the antiadiabatic limit $\omega_0\rightarrow \infty$ where
the effective Hamiltonian Eq. (3) is valid. We will use $\omega_0=1$ here.

Figure 1 shows the effective interaction versus the on-site repulsion $U$. For
$\omega_0=1$ it is attractive for $U<5.9$, while in the antiadiabatic limit it
is attractive only for $U<3.6$. 

It is interesting to consider the structure of energy levels, shown in Figure 2.
We show the energy levels for a single hole in the system and for two holes
in the cases $U=8$ and $U=0$. The effective interaction for these two cases
is $U_{eff}=0.056$ and $U_{eff}=-0.715$ respectively.
There is a clear separation between low-lying energy levels, described by
an effective 'intraband' Hamiltonian and higher-lying levels where the 
spin degrees of freedom are in excited states. In the single particle case
there are 4 low-lying states, in the two-particle case there are 16 low-lying states
for $U=0$ , for $U=8$ there are only 12 because 4 'intraband' states are pushed
high in energy due to the strong on-site repulsion. Note also that the energy 
range of the low-lying states is considerably larger for the case $U=0$ than
for $U=8$. This reflects the bandwidth expansion that occurs upon pairing
in this model.

Figure 3 shows the frequency-dependent conductivity for on-site repulsion $U=8$ (solid
line) and for $U=0$ (dashed line). For $U=8$ the holes are not bound since
$U_{eff}>0$, while for $U=0$ the holes are bound with $U_{eff}=-0.715$. The
low frequency 'intraband' conductivity is seen to increase substantially when
the holes are paired. Furthermore, the optical absorption at intermediate
frequencies ($\omega_0\sim 6$) increases upon pairing, while the
optical absorption at the highest frequencies ($\omega>10$) decreases upon
pairing. This shows that optical spectral weight is transfered from high to
low frequencies when pairing occurs. Interestingly, not only the 'intraband' 
optical spectral weight increases but also spectral weight at intermediate 
frequencies. This will be discussed further in the next section. Note that changes in
optical spectral weight occur at energies much higher than the scale
of the pairing energy, $|U_{eff}| \sim 0.7$.

In Figure 4a we show the optical spectral weight associated with the 
intra-band optical transitions $A_l$ as well as the total optical spectral weight
$A_l+A_h$ versus on-site repulsion $U$. As mentioned earlier the
model does not conserve total optical spectral weight, hence
$A_l+A_h$ is not a constant; nevertheless, the intraband
spectral weight increases faster than the total spectral weight as pairing occurs,
and the ratio $A_l/A_h$ increases as $U$ decreases as shown in Figure 4b.
Hence the model does describe a transfer of optical spectral weight from high
frequencies to low frequencies as pairing occurs.

Figure 4a also shows the optical spectral weight in the antiadiabatic limit
$\omega_0\rightarrow \infty$. As $\omega_0$ increases the high frequency optical
spectral weight moves to higher frequencies and decreases in amplitude, and
vanishes for $\omega_0 = \infty$. However the low frequency $\sigma_1(\omega)$ shows
very little dependence on $\omega_0$ and its total weight is almost the same for
$\omega_0=1$ and $\omega_0=\infty$ as seen in Figure 4a.

The 'undressing' process that occurs upon pairing is also clearly seen in the one-particle
spectral functions. Figure 5 shows the spectral functions for single hole destruction
in the system with one hole,
$A_{10}(\omega)$, and in the system with two holes, $A_{21}(\omega)$. The lowest
frequency peaks are actually $\delta$-functions at zero energy, their weight is the 
quasiparticle weight. For the system with a single hole, the quasiparticle weight
is $z=0.174$. This is larger than the quasiparticle weight for a single site for
these parameters ($z=0.1$); in the extended system, finite $\omega_0$ gives rise to a
larger quasiparticle weight due to retardation. The quasiparticle weight for the
two-hole system and large on-site $U$ is very similar to the single hole spectral
weight, $z=0.176$. When $U$ decreases and a pair is formed, the quasiparticle weight
increases, to $z=0.422$ when $U=0$. As seen in Figure 5b, spectral weight is transfered
from the incoherent region of the spectral function at energies around $\omega \sim 6$ to the
quasiparticle peak. This
energy range corresponds to states where one of the background spins in the system
is in a final excited state. There is also a very small but finite spectral weight in the energy
region around $\omega\sim 12$, where two of the background spins are in excited states
in the final state.

Figure 6 shows the spectral function for hole creation in a system of zero and one hole,
$A_{01}(\omega)$ and $A_{12}(\omega)$. Unlike the spectral functions for hole
destruction they satisfy the sum rule
\beq
\int_0^\infty d\omega A_{n,n+1}(\omega)=1
\eeq
($n=0$ or $1$) because they have no negative frequency component. The quasiparticle
weights extracted from these spectral functions are of course the same as those
obtained from Figure 5. It is interesting to
note that in the intra-band energy range ($\omega < 2$) 
there is now an incoherent contribution to the
spectral function $A_{12}(\omega)$ for parameters where the effective interaction is
attractive. The 'intraband' part of the ground state pair wavefunction is of the form
\beq
|\Psi>=\sum_k f(k)c_{k\uparrow}^\dagger c_{-k\downarrow}^\dagger |0>
\eeq
and it has finite overlap with states other than $c_{0\uparrow}^\dagger c_{0\downarrow}^\dagger |0>$.
Still the weight of the coherent part of the spectral function, i.e. the
zero frequency peak 
(quasiparticle weight) increases strongly as the pair is formed.

The behavior of quasiparticle weights as function of on-site repulsion is shown in Figure 7.
For large repulsive $U$, the quasiparticle weight for a hole is the same for the
system with two holes (dashed line) and with one hole (full line); as the on-site
repulsion decreases and the pair forms, the quasiparticle weight in the system
with two holes increasing, indicating that 'undressing' occurs. We also show the
corresponding results in the $\omega=\infty$ limit, which show similar behavior; the
quasiparticle weights in this case are smaller than for finite frequency. For the
single hole, the quasiparticle weigth in the antiadiabatic limit is the same
as for the single site, $z=S^2$.

In summary, these results show that in the dynamic Hubbard model there is
'undressing' when pairing occurs: spectral weight 
 in both one- and two-particle spectral functions is transfered from high to
low frequencies, the quasiparticle weight increases and the kinetic energy
decreases when
pairing occurs. We next discuss the situation for the electron-hole symmetric model.

\subsection{Electron-hole symmetric Holstein-like model}

We consider the electron-hole symmetric (e-h sym) model Eq. (5), for
parameters $g=3$ and $\omega_0=1$. The site energies for $n$ holes in this case are
\bmath
\beq
\epsilon(n)=-\epsilon(\bar{n})=-\omega_0\sqrt{1+g^2}
\eeq
for $n=0$ and $n=2$, and
\beq
\epsilon(n)=-\epsilon(\bar{n})=-\omega_0
\eeq
\emath
for $n=1$.
Because the excitation energy of the background spin is lower
for the singly occupied site here there is not such a clear separation of 
energy scales for this model as for the dynamic Hubbard model for the parameters
used.

Figure 8 shows the effective interaction between
two holes in the four-site cluster versus on-site repulsion, as well as the results
in the antiadiabatic limit $\omega_0=\infty$. Note that retardation is detrimental
to pairing in this model. For $\omega_0=1$, the effective interaction is attractive
for on-site repulsion $U$ smaller than $\sim -0.4$.

The optical conductivity for two holes in the cluster is shown in Figure 9,
for on-site repulsion $U=2$, where the effective interaction is repulsive, and
for $U=-2$ where it is attractive. In this case a transfer of optical spectral
weight from low to high frequencies occurs upon pairing, opposite to the
behavior in the dynamic Hubbard model seen in Figure 3. Note also that the
separation between low and high energy regions is less clear here than in the
previous case, as expected.

The dependence of optical spectral weights on the on-site repulsion is shown in Figure 10a.
The intra-band spectral weight decreases by a factor of 2 as pairs form; this indicates
that the carriers are more heavily dressed and have larger effective mass in the
paired state, as one would expect; in the $\omega\rightarrow \infty$ limit this
model becomes equivalent to an attractive Hubbard model, where the pair mobility
is always smaller than the single particle mobility\cite{bose}; in particular in the strong
coupling limit the pair hopping is $t_{p}=2t^2/|U|$ much smaller than the
single particle hopping for large on-site attraction. This model then describes a
transfer of optical spectral weight from low to high frequencies when pairing
occurs; the ratio of intra-band to inter-band optical spectral weights versus
on-site $U$ is shown in Fig. 10b, qualitatively different from the corresponding
results for the dynamic Hubbard model Figure 4b.

The single particle spectral functions for hole destruction are shown in Figure 11.
The quasiparticle weight for the system with a single hole is $z=0.89$, for
the system with two holes it is $z=0.79$ for the case $U=2$ with unpaired holes,
and it decreases to $z=0.64$ for $U=-2$ when the holes are paired. Note that 
this is not a single-site effect, as for a single site the spectral weight for
one and two holes is identical due to electron-hole symmetry (Eq. (19)).
Figure 11b shows that spectral weight from the quasiparticle peak is transfered
to higher frequencies around $\omega\sim 3$ when pairs form. Similarly, Figure 12
shows the single particle spectral function for hole creation in the system
with zero and with one hole. For the creation of a hole in the empty cluster the
incoherent part is here at lower frequencies due to the lower excitation energy of
the singly-occupied site, Eq. (24b). Again, figure 12b shows that as pairing
occurs spectral weight is transfered from the quasiparticle peak to higher frequencies.

Figure 13 shows the variation of quasiparticle weight versus on-site repulsion
in this model.
Because for the single site the quasiparticle weight is
independent of hole occupation the variation here is less than in the dynamic
Hubbard model (Fig. 7). The quasiparticle weight in the system with 2 holes
decreases as the pair formed, indicating that the quasiparticle is more
heavily dressed in the paired compared to the unpaired state, qualitatively
different to the situation in the dynamic Hubbard model.

Figure 13 also shows the quasiparticle weights in this model in the antiadiabatic
limit, where it is equivalent to the Hubbard model. Note that the quasiparticle
weights for the system with 1 and 2 holes coincide for $U=0$, where the 
Hamiltonian describes non-interacting holes, and are both given by
the site value $z=S^2=0.658$; both for repulsive and attractive $U$ the quasiparticle weight
is smaller in the two-hole system due to 'intra-band' hole-hole interaction.
The suppression of quasiparticle weight however is larger for negative than
for positive $U$.

\subsection{Summary}

The qualitatively different behavior of the two models considered is
summarized in Figure 14. We plot the effective mass enhancement and the
quasiparticle weight as function of the effective interaction $U_{eff}$ in both
models. The ratio of effective masses for the single particle and the
pair, $m_s^*/m_p^*$, is obtained from the ratio of the intra-band
optical spectral spectral weight for the system with two holes and
twice the intra-band optical spectral weight for the system with one hole.
Figure 14a shows that in the dynamic Hubbard model 
the pair becomes increasingly lighter than the single
particle as the interaction becomes more attractive, while in the electron-hole
symmetric model the pair becomes increasingly heavier as the interaction becomes
more attractive. Similarly, the quasiparticle weight increases in the dynamic
Hubbard model as the pair is formed, while it decreases in the electron-hole
symmetric model. In summary, the system becomes more coherent in the paired
state in the dynamic Hubbard model and more incoherent in the electron-hole
symmetric model.

It is interesting to note that the quasiparticle weight and the effective mass
change by approximately the same factor in the dynamic Hubbard model in the parameter
range considered, 2.5 and 2.4 respectively. This is what one would expect if the physics
of the model is dominated by 'single site' physics, where the quasiparticle weight 
and effective hopping and hence effective mass are
both determined by the single site overlap matrix element $S$. More generally, in a many body system the
exact single particle Green's function is given by\cite{mahan}
\beq
G(k,\omega)=\frac{1}{w-\epsilon_k-\Sigma(k,\omega)}=\frac{z_k}{\omega-\tilde{\epsilon}_k}+G'(k,\omega)
\eeq
with $\Sigma(k,\omega)$ the self-energy, $\tilde{\epsilon}_k$ the quasiparticle energy
and $G'$ the incoherent part of the Green's function. The quasiparticle weight 
$z_k$ and effective mass enhancement are given by
\bmath
\beq
z_k=(1-\frac{\partial}{\partial \omega}\Sigma_{re}(k,\omega))^{-1}
\eeq
\beq
\frac{m}{m^*}=\frac{\partial \tilde{\epsilon}_k}{\partial \epsilon_k}=
z_k(1+\frac{\partial \Sigma_{re}(k,\omega)}{\partial \epsilon_k})
\eeq \emath
($\Sigma_{re}=$ real part of $\Sigma$) so that if the self-energy is  
momentum independent the
quasiparticle weight and effective mass renormalization coincide.
Hence our results indicate that this is approximately the situation in 
the dynamic Hubbard model, and suggest that dynamical mean field theory\cite{jarrell},
which assumes a momentum-independent self-energy, should be a useful
approach to study these models. In contrast, 
the results for the electron-hole symmetric model in the parameter range considered
yield an effective mass changing by a factor 1.9 with the quasiparticle weight 
changing by a factor of only 1.2 , suggesting that for this model the momentum-dependence
of the self-energy is substantial.

\subsection{Finite temperatures}

We have also studied the behavior of the optical conductivity in the
dynamic Hubbard model at finite temperatures, given by
\beqn
\sigma_1(\omega)=\frac{\pi}{Z}\sum_{n,m}\frac{e^{-\beta E_n}-e^{-\beta E_m}}{E_m-E_n}
&|&<0|J|m>|^2 \times \\ \nonumber
&\delta &(\omega-(E_m-E_n))
\eeqn
with $Z=\sum_n e^{-\beta E_n}$ the partition function. Figure 15 shows $\sigma_1(\omega)$ 
for a case where the effective interaction is attractive, $U_{eff}=-0.715$. The 
intra-band part of the optical absorption is rapidly suppressed as the
temperature increases, and the peak at intermediate frequencies is also
suppressed. Unfortunately because tight binding models do not satisfy the optical
sum rule the total optical spectral weight in the model is not conserved as
the temperature changes. Nevertheless it is interesting to note that as
$T$ increases optical spectral weight is transfered to the very high
frequency region $\omega >10$. This change in high frequency spectral
weight occurs on a temperature scale ($T\sim 1$) that is related to the
energy scale of the pairing energy ($U_{eff}$) and unrelated to the
energy scale of the frequency where the optical absorption change occurs.

\section{Relation with experiments}

The dynamic Hubbard model describes a coupling of electrons  to a background degree
of freedom (the auxiliary spin) that only exists when the site is doubly occupied by electrons.
Hence its effect decreases with increasing local hole concentration, on the
average the coupling constant is 
\beq
\lambda(n)=g\omega_0 (2-n)
\eeq
with $n$ the average hole concentration per site. This leads to a phenomenology whereby
hole carriers 'undress' in the presence of other hole carriers. The undressing
manifests itself in transfer of optical spectral weight from high to
low frequencies and in transfer of one-particle spectral weight from the incoherent
(high frequency) to the coherent (low frequency) region; these spectral weight
transfers lead to increase in the quasiparticle weight and decrease in the
quasiparticle effective mass. Because a hole comes close to another hole both
when the hole concentration is increased by doping in the normal state as well
as when carriers pair, undressing will occur both for increasing carrier concentration
and for decreasing temperature.

There is substantial evidence for such phenomenology in the high $T_c$ cuprates.
Johnson et al\cite{johnson} extract from photoemission experiments in the normal state a
coupling constant that decreases continuously as the hole doping increases,
as well as a mass enhancement that decreases with hole doping. In the overdoped
regime, Yusof et al\cite{yusof} find evidence from photoemission for the
existence of quasiparticles in the normal state, which appear to be absent in
the underdoped regime. This is consistent with the phenomenology of the
dynamic Hubbard model in a strong coupling regime where the quasiparticle weight
for low hole density would be small enough to be unobservable. The quasiparticle
weight in the dynamic Hubbard model as function of hole concentration is
approximately given by
\bmath
\beq
z(n)=S^2(1+\frac{n}{2}\Upsilon )^2
\eeq
\beq
\Upsilon =\frac{1}{S}-1
\eeq
\emath
If the coupling strength $g$ is large $S$ will be small, $z$ becomes very small in 
the underdoped regime, and at the same time the 'undressing parameter' $\Upsilon$ becomes
large, hence quasiparticles undress rapidly with increasing $n$.

Ando et al\cite{ando} extract from transport measurements a hole mobility that
increases monotonically with hole doping, consistent with the behavior predicted
by the dynamic Hubbard model and its low energy effective Hamiltonian Eq. (3), which
leads to an effective density-dependent hole hopping
\beq
t_{eff}(n)=t_2+n\Delta t
\eeq
which is equivalent for low $n$ to the more fundamental relation
\beq
t_{eff}(n)=t z(n)
\eeq
describing the fact that the quasiparticle effective mass is inversely proportional
to the quasiparticle weight as expressed by Eq. (26b) when the self-energy has
no momentum dependence.

Ding et al\cite{ding} make a compelling description of the undressing phenomenology
of high $T_c$ cuprates: from their photoemission data they extract a quasiparticle
weight that emerges from an incoherent background as the temperature is lowered
and the system becomes superconducting. This is of course consistent with the
phenomenology of the dynamic Hubbard model in the strong coupling regime where
$z$ will increase strongly when pairing occurs, as seen in Figs. 5 and 7. Furthermore
Ding et al find that $z$ increases as the doping increases, again consistent with the
behavior expected in the dynamic Hubbard model.

In the finite cluster calculations reported here for the optical conductivity
we cannot distinguish whether the optical spectral weight transfered to low
frequencies when pairing occurs goes into the zero-frequency $\delta-$function
or to finite frequencies. However the effective Hamiltonian for the dynamic Hubbard
model clearly describes transfer of spectral weight to the $\delta-$function as
the pairing amplitude develops\cite{apparent,london}, since the average
effective kinetic energy
\beqn
T_{eff}=& &-\sum_{i,\sigma}[t_{eff}(n)<c_{i\sigma}^\dagger c_{i+1\sigma}+h.c.)> \\ \nonumber
& &-2\Delta t (<c_{i\sigma}^\dagger c_{i\-\sigma}^\dagger> <c_{i\-\sigma} c_{i+1\sigma}>+h.c.]
\eeqn
has a contribution from  anomalous expectation values, while the intra-band optical
absorption is unchanged except for the depletion due to the opening of the
superconducting energy gap\cite{coherence}. This indicates that the optical
spectral weight transfered into the intra-band region goes into the
zero-frequency $\delta-$function, and will lead to an apparent
violation of the Ferrell-Glover-Tinkham optical sum rule\cite{apparent}.
This violation has been observed experimentally, both for in-plane as well as
c-axis  light polarization by Santander-Cyro et al\cite{santander} and by
Basov et al\cite{basov} respectively. The decrease
in high frequency spectral weight predicted by the dynamic Hubbard model
 (range above $\omega \sim 10$ in Figs. 3 and 15) is also consistent with recent
experimental observations by Molegraaf et al\cite{marel}, who report a
decrease in optical spectral weight in the energy range between
$1.25eV$ and $2.5eV$ which is transfered to lower frequencies. This is also
consistent with earlier observations by Fugol et al\cite{fugol}.

Note that our results for the dynamic Hubbard model predict also an increase in
optical spectral weight at intermediate frequencies, $\omega \sim 6$ 
when the interaction becomes attractive (Fig. 3) or when the temperature
decreases (Fig. 15),
well above the 'intra-band' frequency range. We expect this to be a generic
feature of these models, describing optical transitions where a hole is
transfered to a nearest neighbor site already occupied by another hole,
leaving the auxiliary spin behind in an excited state. Such a 
'vertical' transition will have a large weight in the
paired state (proportional to $1$ rather than to $S$ if the neighboring site
is unoccupied) and enhances the optical absorption at frequencies corresponding
to the excitation energy of a single site. Given the correspondence discussed
above between the range $\omega>10$ in our example and the visible frequency
range, the intermediate region $\omega \sim 6$ would correspond to the
mid-infrared range of frequencies in the cuprates. Indeed, Gao et al\cite{gao}
report observation of extra optical spectral weight appearing below $T_c$ in
the mid-infrared region.

Table I summarizes the matrix elements for the various optical transitions in
the dynamic Hubbard model considered here for the case of a single hole
versus the case of two holes on neighboring sites. The weight ratios given
in the last column summarize the expected qualitative behavior of spectral
weight changes when hole doping increases in the normal state or when
the temperature is lowered and the system goes superconducting. We expect the 
frequency range of the
second and third rows to correspond to mid-infrared and that of the
fourth row to visible frequencies in the cuprates. Because
$S+1/S >2S$ for any $S$, the optical spectral weight in the 
mid-infrared range should always increase
upon pairing.

\section{Discussion}
We have studied in this paper the behavior of spectral functions in a dynamic
Hubbard model by exact diagonalization. The results obtained are exact for
the small cluster studied. They support a scenario for the physics of this
class of quantum many-body systems that is expected from qualitative 
arguments and approximate treatments. Namely, that quasiparticles 'undress'
in these models when the local hole concentration increases, which occurs both
when holes are added to the system (doping) and when holes pair and
form Cooper pairs. This scenario is obtained from the exact calculations in this
paper without uncontrolled approximations.

The physics of dynamic Hubbard models is especially transparent in the 
antiadiabatic limit, where the effective Hamiltonian is a Hubbard model
with correlated hopping. The results of ref. \cite{qmc} as well as the
results in this paper indicate that the 'intra-band' physics of the model
described by the antiadiabatic limit $\omega_0\rightarrow \infty$ remains
essentially unchanged with $\omega_0$ decreasing to rather small values.
In addition to the case $\omega_0=1$ discussed in this paper we also studied the
model for $\omega_0=0.5$ and obtained qualitatively similar results. What changes
as $\omega_0$ decreases is that the energy scale of non-intraband excitations
decreases up to a point where there is no longer a clear separation between
'intra-band' and 'non-intraband' regions. Nevertheless the low energy physics
remains essentially unchanged.

In a more realistic description of a real system there will presumably be a set of
excitation energies $\omega_i$ describing the bosonic excitations that dress
the hole quasiparticles. Still we do not expect the physics in such a case to be
qualitatively different. These excitations would correspond to local electronic
excitations with scale up to eV's. Such energy scales are consistent with 
experimental observations in cuprates that optical spectral weight in the visible
range is transfered to low frequencies both when the system goes superconducting\cite{marel}
as well as when it is hole-doped in the normal state\cite{uchida}. Our results in
this paper as well as earlier results\cite{color} demonstrate that dynamic
Hubbard models naturally describe transfer of spectral weight from high energies
unrelated to the scale of superconductivity down to low frequencies when superconductivity
sets in, as observed\cite{marel,fugol}, hence they provide a natural explanation
for the origin of the high energy scale observed in the
optical experiments of Molegraaf et al\cite{marel}. 
The detailed description of 
such apparently  counterintuitive physics remains a challenge for other proposed
descriptions of the physics of high temperature superconductors that also propose that
superconductivity is driven by kinetic energy lowering\cite{kiv,and}.

The dynamic Hubbard model and the electron-hole symmetric model
discussed in this paper are representative of two classes of
model Hamiltonians, of which there are many different realizations.
In particular, it is not essential that the coupling of the boson in a model in the class
of dynamic Hubbard models be to the on-site double occupancy; a model with
coupling only to the on-site charge density will also belong to this class
if it is not electron-hole symmetric\cite{hole1}. Also the auxiliary boson
may be an oscillator rather than a spin, or the model could have only
electronic degrees of freedom\cite{dynh2}. What distinguishes these
two classes of models is what is the driving energetics for pairing:
in dynamic Hubbard models pairing is kinetic energy driven, and the
potential energy increases upon pairing, and the opposite is true in the
other class of models, which may be termed 'conventional' or 'electron-hole symmetric'.
The conventional electron-phonon models used to describe conventional superconductors
belong to this second class of models. We believe that these two classes of models represent
very general paradigms. In the class of dynamic Hubbard models the 'undressing'
of quasiparticles is essential to lead to kinetic energy lowering; instead, in the
'conventional' class of models the dressing of the quasiparticles may remain unchanged upon
pairing if the
coupling is weak, or increase in a strong coupling regime.

Dynamic Hubbard models and their low energy effective Hamiltonians can describe
superconductivity over the entire range of coupling strengths. The physics is
determined by the scale of excitation energies $\omega_0$ and by the strength of
the couping $g$ or equivalently the magnitude of the 'undressing parameter'
$\Upsilon$ (Eq. (28b). As we have seen in this paper and in ref. \cite{qmc} the
low energy physics is not strongly dependent on the scale $\omega_0$. The magnitude
of $T_c$ and the superconducting gap is mainly determined by the strength of the
dimensionless parameters $g$ or $\Upsilon$ and the single electron hopping
parameter $t$, as well of course as competing Coulomb repulsions such as $U$.
These parameters cannot however be tuned separately at will, in a real system
they are all closely interdependent and determined principally by the
ionic charge $Z$ as discussed in refs. \cite{dynh1,dynh2}. For increasing $g$,
$T_c$ becomes large, the coherence length in the superconducting state becomes short
and the system becomes incoherent in the normal state for low hole concentration
as the magnitude of $S$ and the quasiparticle weight decrease; the 'undressing'
phenomenology becomes particularly apparent in this regime. For small $g$ ,
$T_c$ becomes small, the coherence length can become thousands of lattice spacings
 and the normal
state becomes coherent; in this regime, even though it is still the same
'undressing physics' that drives the transition to superconductivity, the
anomalous spectral weight transfers signaling undressing will become almost
invisible. We suggest that it is possible that this same physical
mechanism can describe the superconducting phenomenology of materials as
distinct as $YBa_2 Cu_3 O_{7-\delta}$, $MgB_2$, and $Al$ \cite{narlikar}.

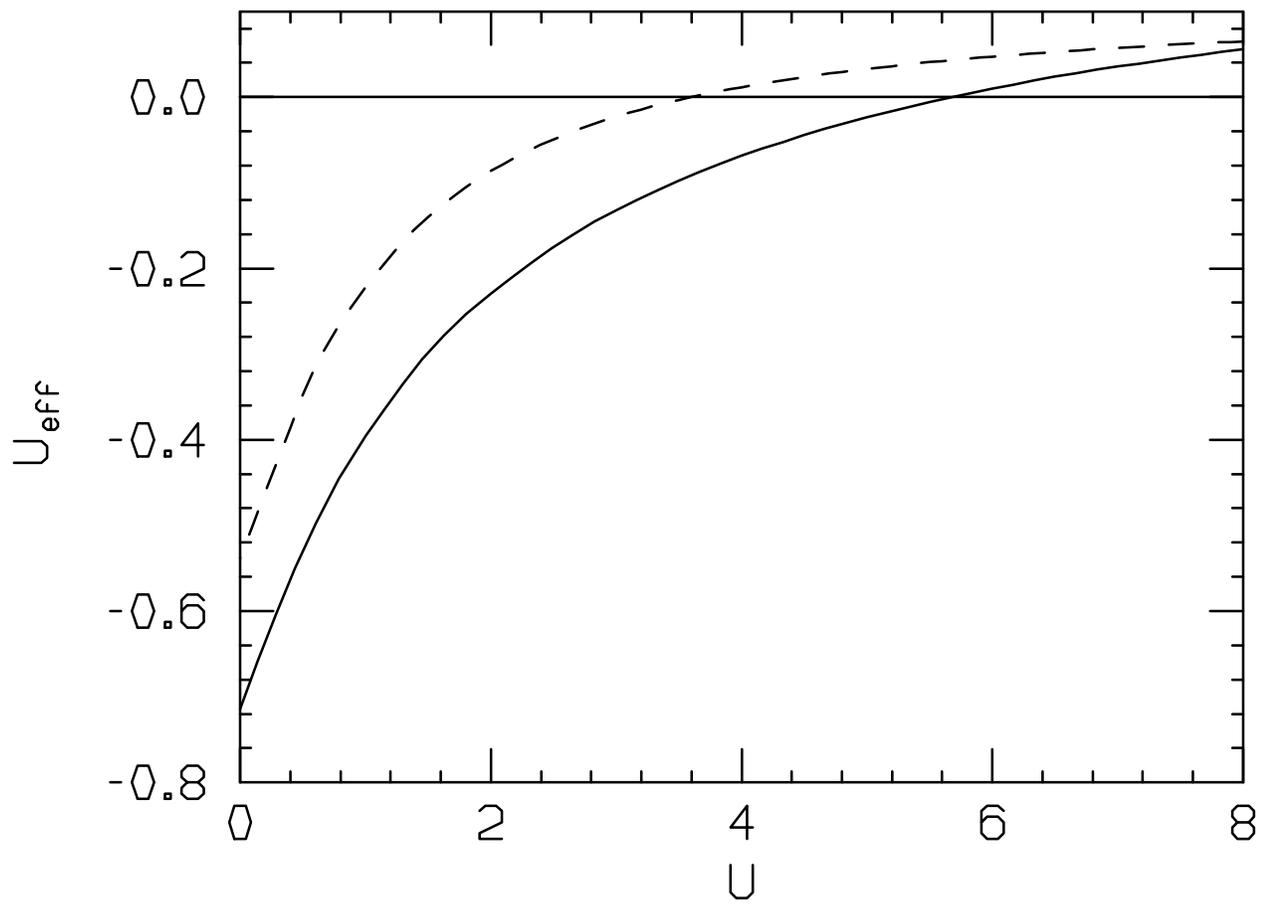
\begin{figure}
\caption { Effective interaction Eq. (21) versus on-site repulsion
$U$ for the dynamic Hubbard model with $g=3$ and $\omega_0=1$ (full line)
and in the antiadiabatic limit $\omega_0=\infty$ (dashed line). The same
parameters ($g=3, \omega_0=1$) are used in the following figures.
}
\label{Fig. 1}
\end{figure}
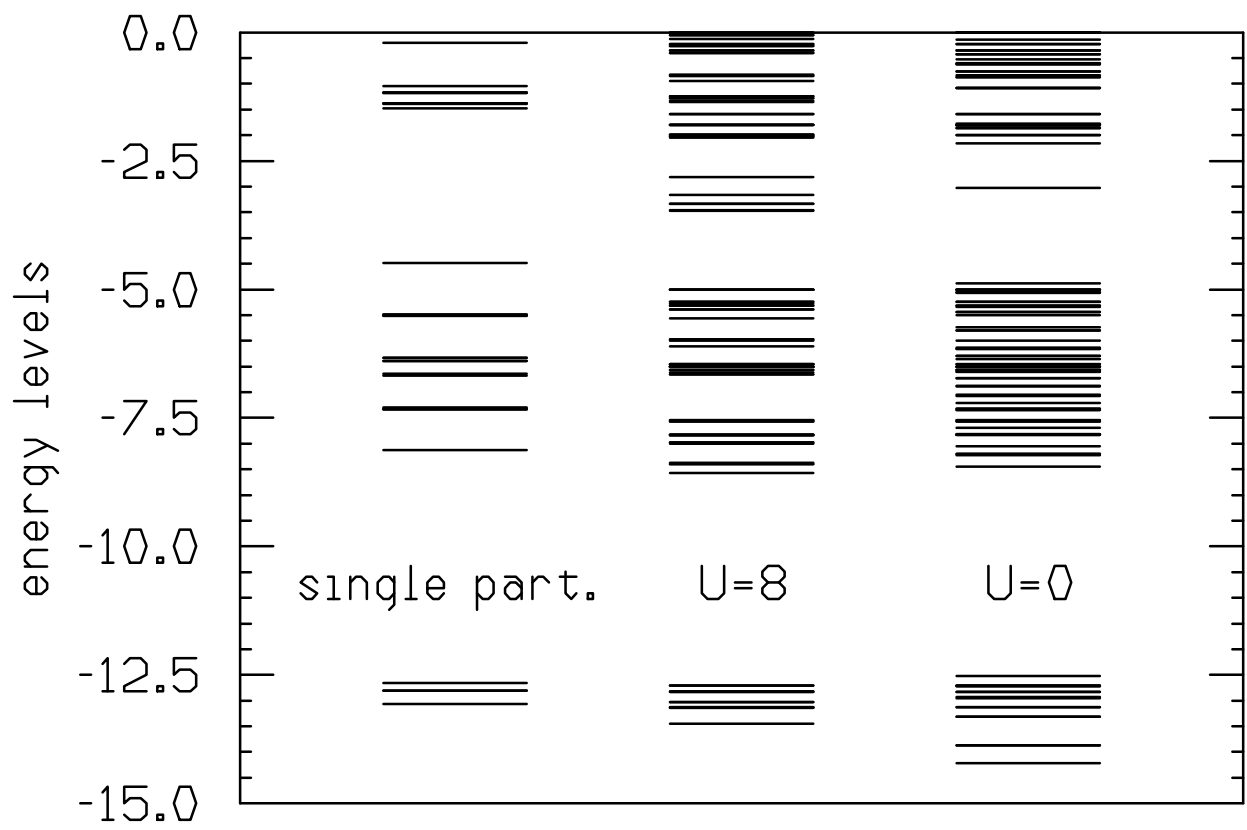
\begin{figure}
\caption {Energy levels for the dynamic Hubbard model. The energy levels for a 
single hole in the four-site cluster and for two holes with on-site repulsion
$U=8$ and $U=0$ are shown.
}
\label{Fig. 2}
\end{figure}
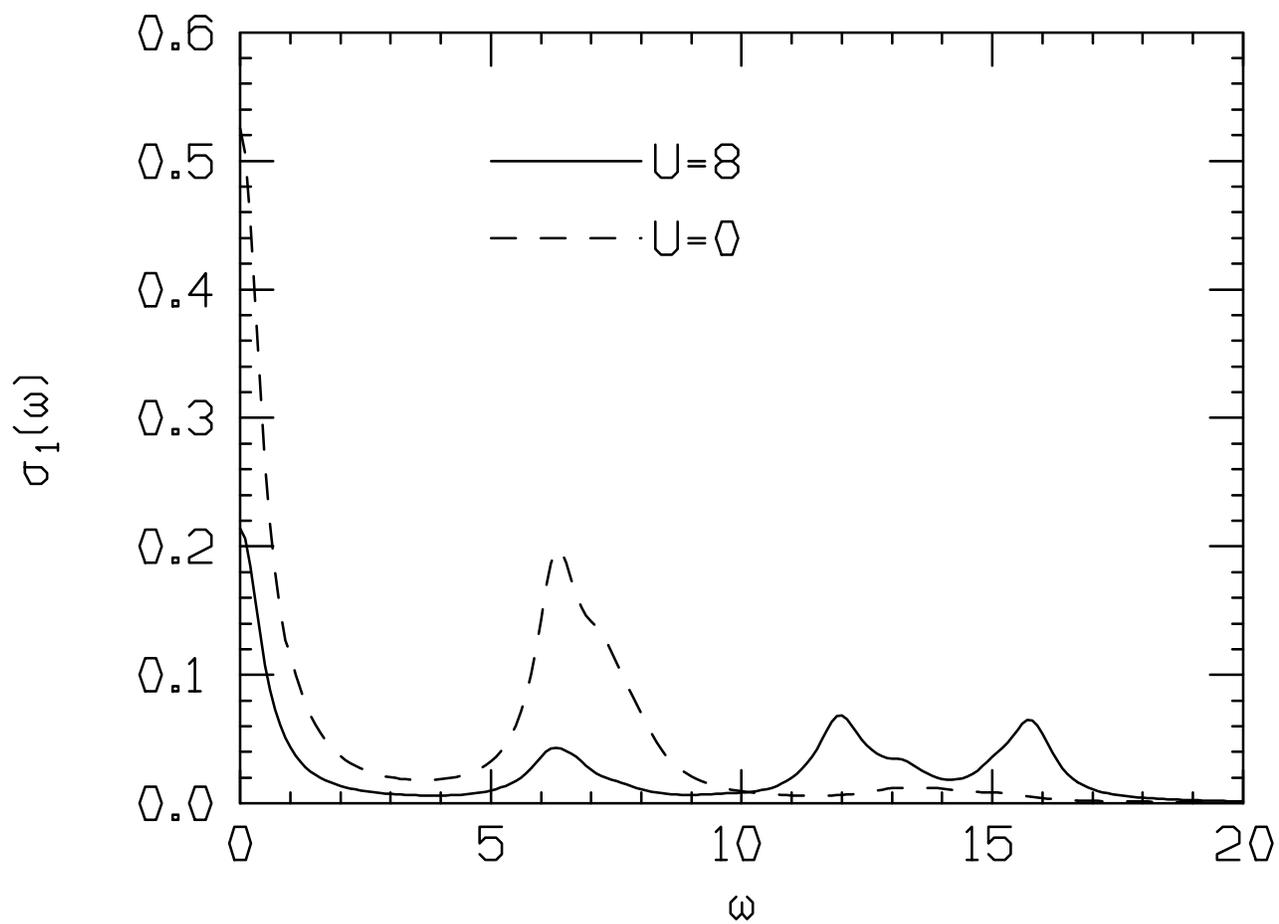
\begin{figure}
\caption {Frequency-dependent conductivity for dynamic Hubbard model with 
two holes and on-site repulsion $U=8$ and $U=0$. The $\delta$-functions in
Eq. (8) are broadened to lorentzians with width $\Gamma=0.5$. The lowest
frequency $\delta-$function at frequency $\omega_D$ ('Drude precursor') is
shifted to $\omega=0$ and represented by a Drude form (semi-lorentzian)
with width $\Gamma=0.5$.
}
\label{Fig. 3}
\end{figure}
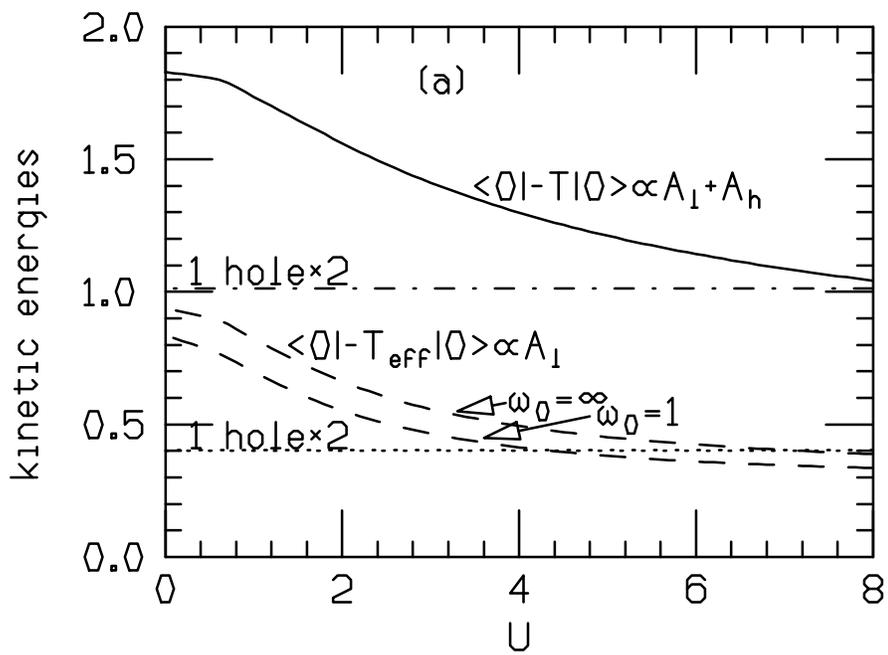
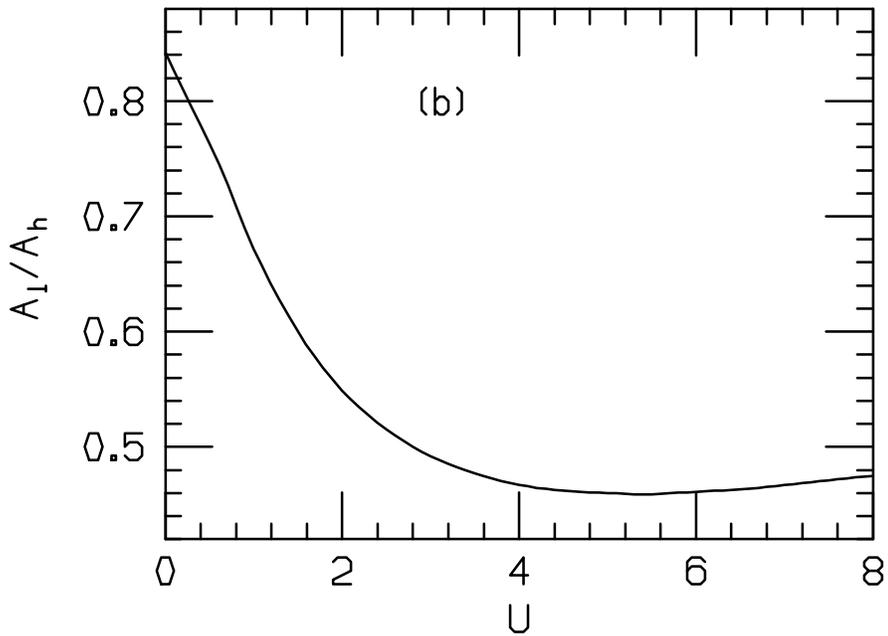
\begin{figure}
\caption {(a) Kinetic energies for dynamic Hubbard model versus on-site $U$, obtained from
integration of $\sigma_1(\omega)$ from Eqs. (12) and (16). The cutoff frequency
to define the 'intraband' spectral weight $A_l$ is $\omega_m=2$. The dash-dotted line
gives the results for twice the total optical spectral weight $A_l+A_H$ for a single
hole in the cluster, which coincides with the results for two holes in the
cluster for large $U$. The dotted line gives twice the intra-band optical spectral
weight for one hole, which is approximately equal to the intra-band spectral
weight for two holes in the cluster for large $U$. The intra-band spectral weights
for two holes are shown both for $\omega_0=1$ and in the antiadiabatic limit
$\omega_0=\infty$; for one hole in the antiadiabatic limit the results are
indistinguishable from the value for $\omega_0=1$. (b) Ratio of
intra-band to inter-band optical spectral weight versus $U$ for $\omega_0=1$.
}
\label{Fig. 4}
\end{figure}
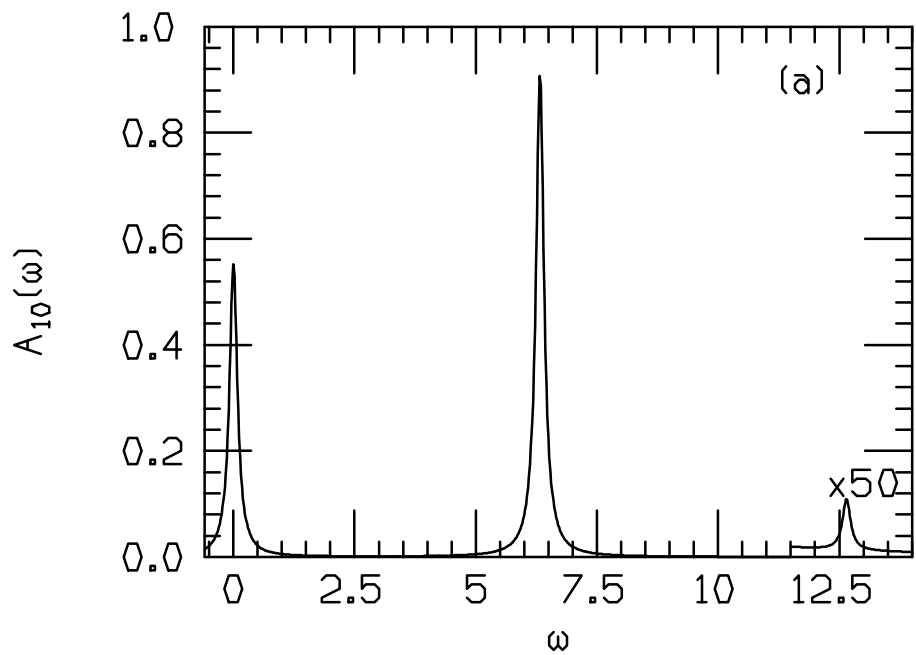
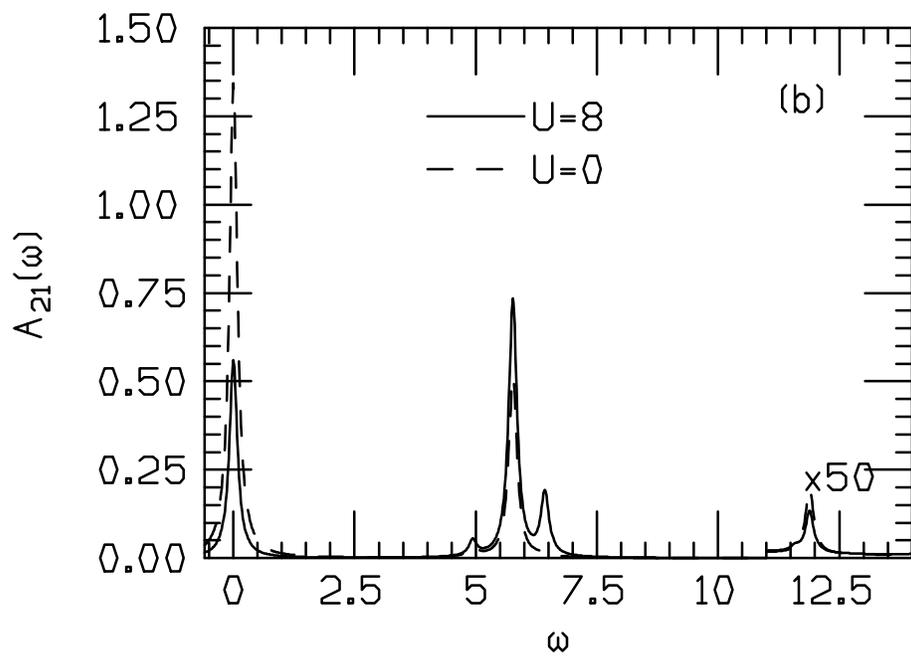
\begin{figure}
\caption {One-particle spectral function for hole destruction in the system
with one hole (a) and with two holes (b). The $\delta-$ functions are
broadened to lorentzians with width $\Gamma=0.1$. The spectral weight at
frequencies above $\omega=10$ is very small and is amplified in the figure
by a factor $50$.
}
\label{Fig. 5}
\end{figure}
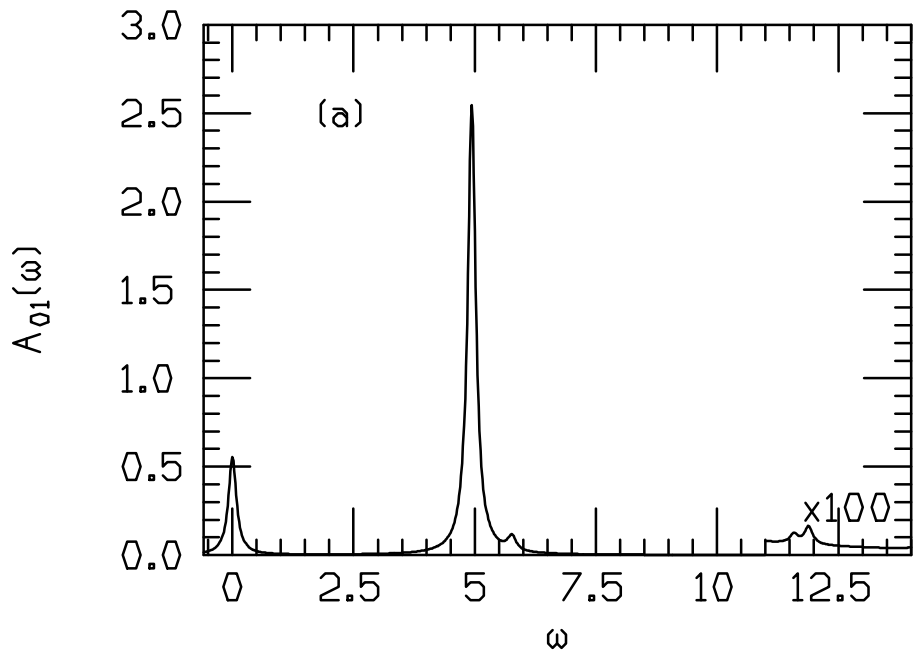
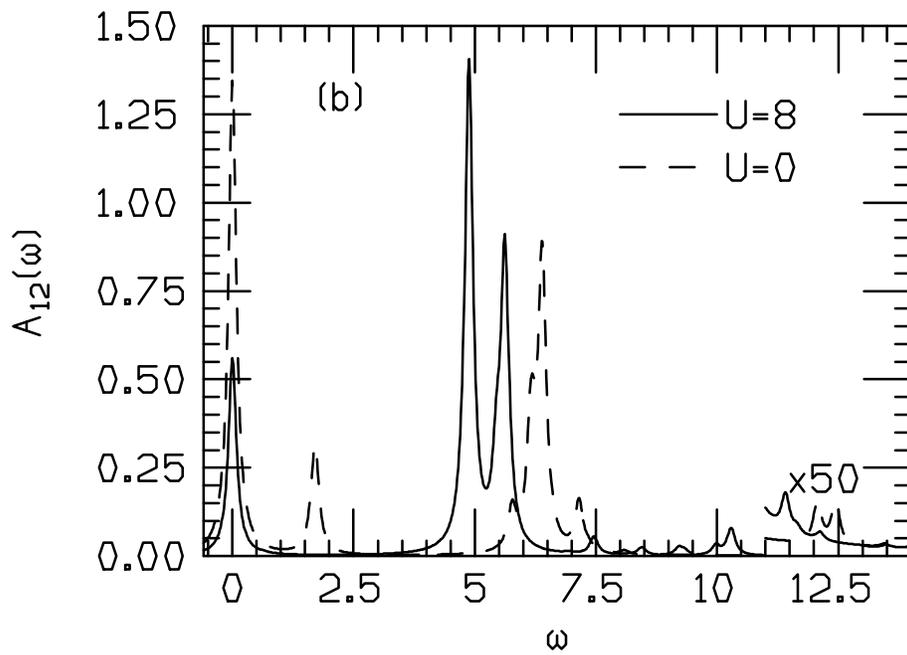
\begin{figure}
\caption {Same as figure 5 for hole creation in the system with zero holes (a) 
and with one hole (b).
}
\label{Fig. 6}
\end{figure}
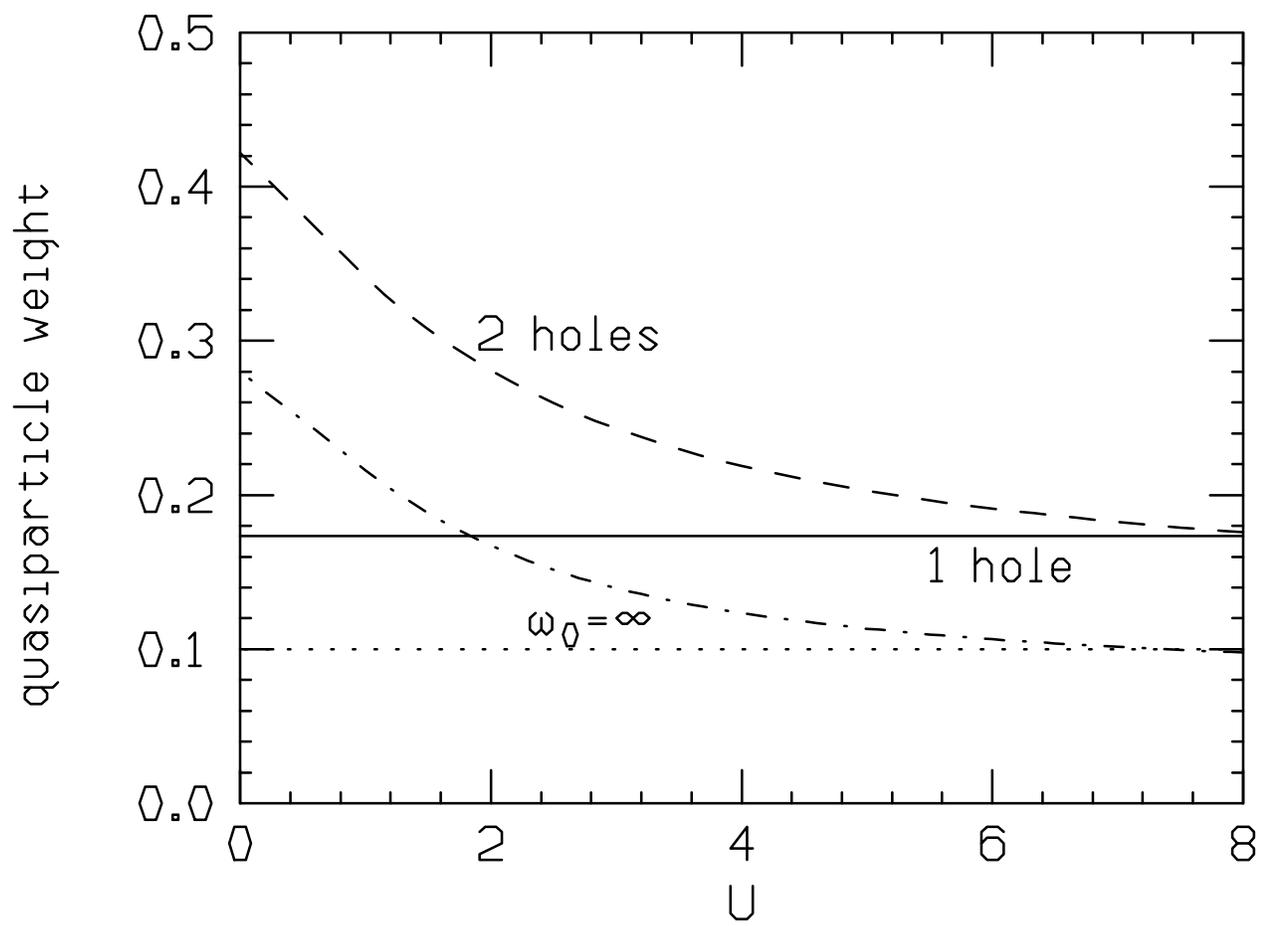
\begin{figure}
\caption {Quasiparticle weights versus $U$ for the system with two holes (dashed line)
and with one hole (full line). The corresponding quasiparticle weights
in the antiadiabatic limit are shown as the dash-dotted and dotted lines
respectively.
}
\label{Fig. 7}
\end{figure}
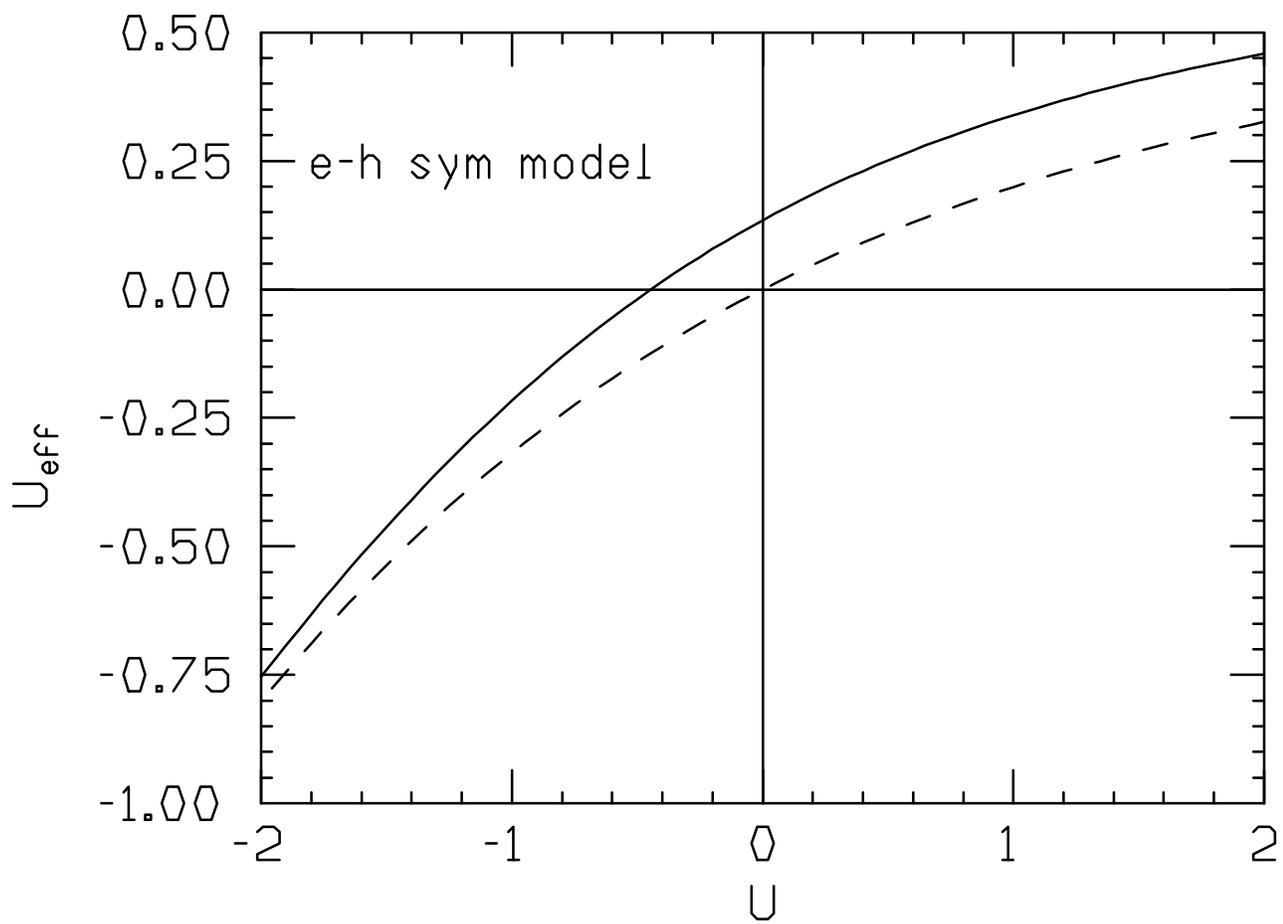
\begin{figure}
\caption { Effective interaction Eq. (21) versus on-site repulsion
$U$ for the electron-hole symmetric model Eq. (5),  with $g=3$ and $\omega_0=1$ (full line)
and in the antiadiabatic limit $\omega_0=\infty$ (dashed line). The same
parameters ($g=3, \omega_0=1$) are used in the following figures.
}
\label{Fig. 8}
\end{figure}
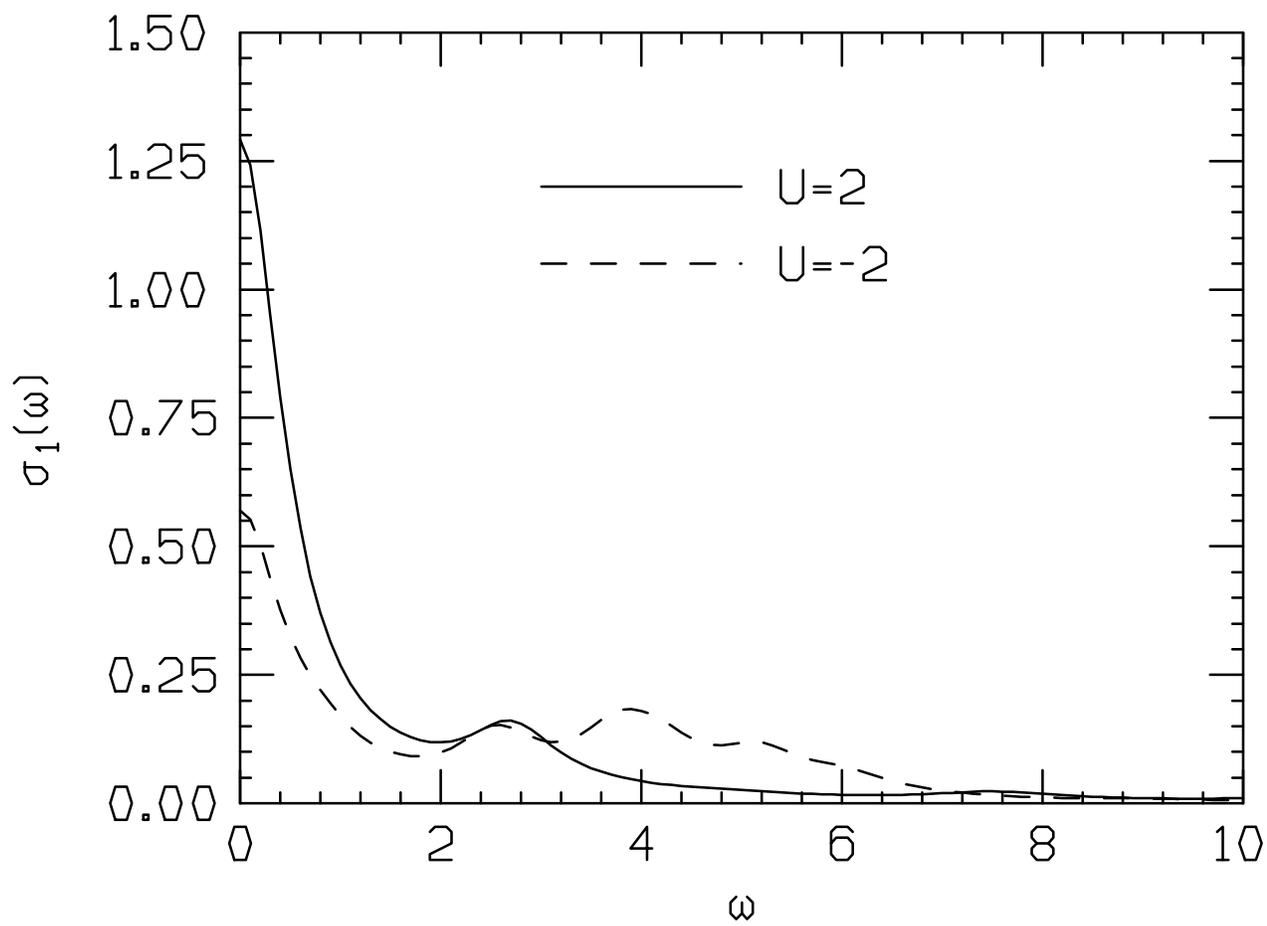
\begin{figure}
\caption {Frequency-dependent conductivity for the electron-hole symmetric
model  with 
two holes and on-site repulsion $U=2$ and $U=-2$. 
}
\label{Fig. 9}
\end{figure}
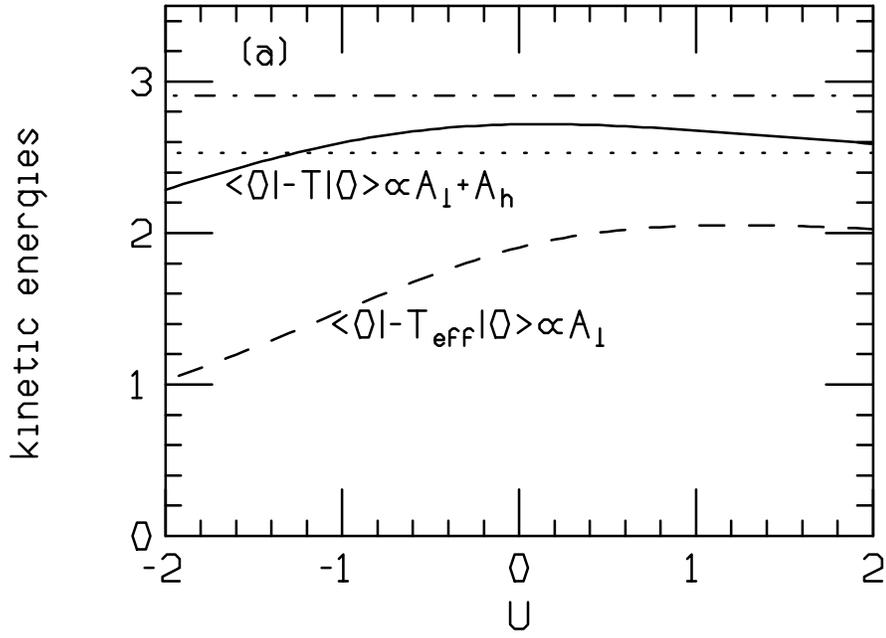
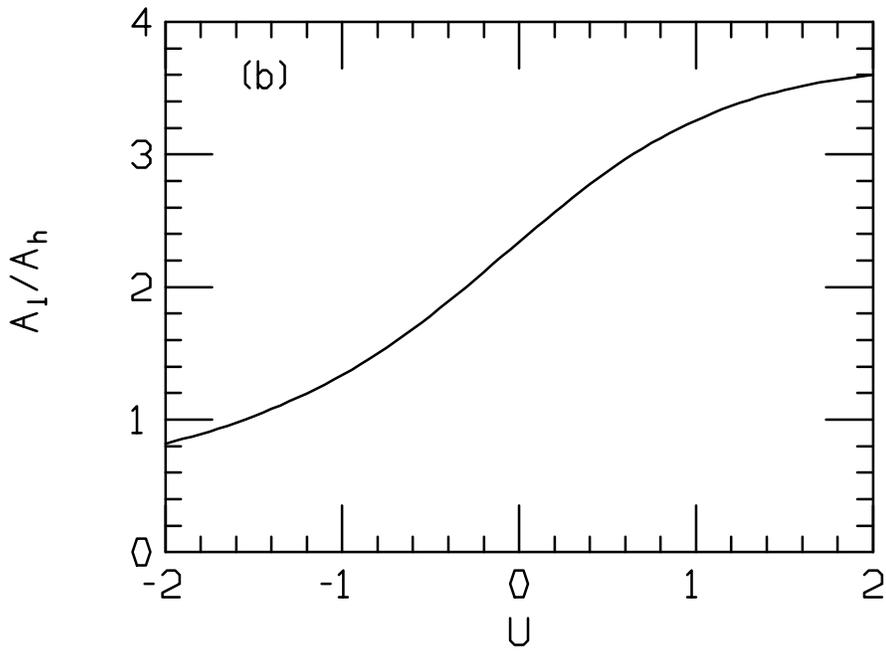
\begin{figure}
\caption {(a) Kinetic energies for electron-hole symmetric model versus on-site $U$. The cutoff frequency
to define the 'intraband' spectral weight $A_l$ is $\omega_m=2$. The dash-dotted line
gives the results for twice the total optical spectral weight $A_l+A_H$ for a single
hole in the cluster, and the dotted line gives the corresponding intra-band value.  (b) Ratio of
intra-band to inter-band optical spectral weight versus $U$.
}
\label{Fig. 10}
\end{figure}
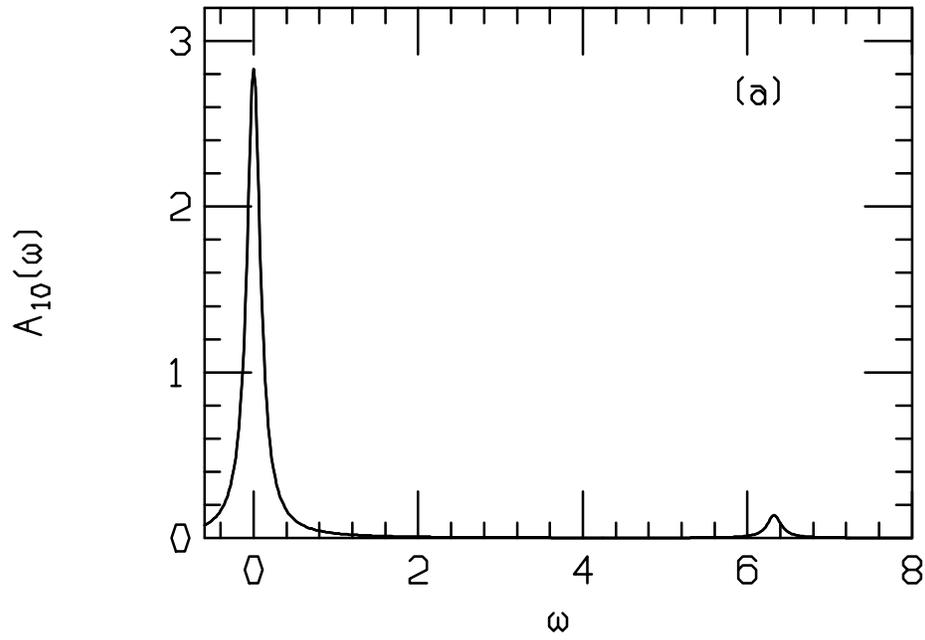
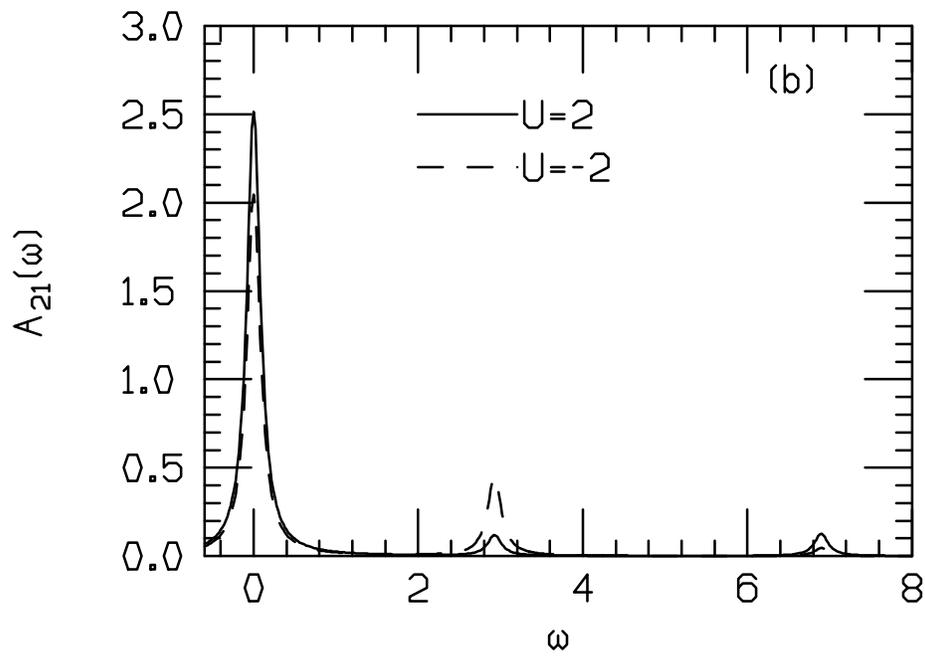
\begin{figure}
\caption {One-particle spectral function for hole destruction in the 
electron-hole symmetric model 
with one hole (a) and with two holes (b).
}
\label{Fig. 11}
\end{figure}
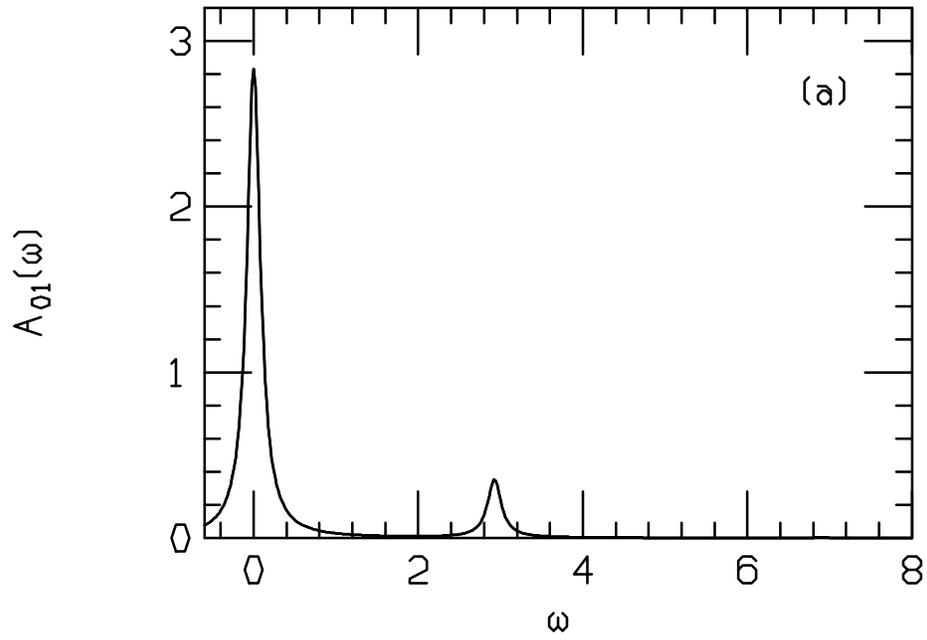
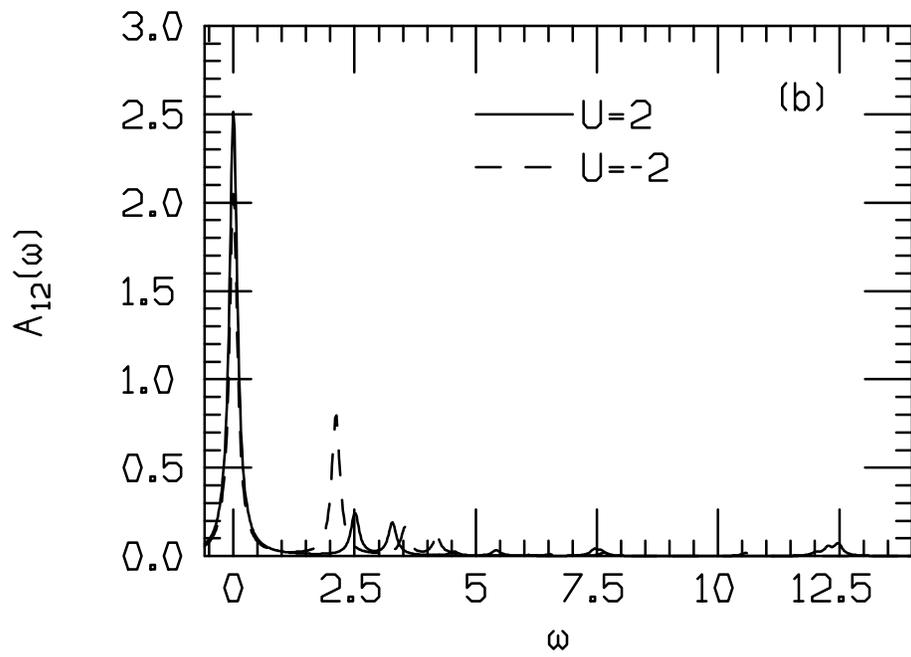
\begin{figure}
\caption {Same as figure 11 for hole creation in the system with zero holes (a) 
and with one hole (b).
}
\label{Fig. 12}
\end{figure}
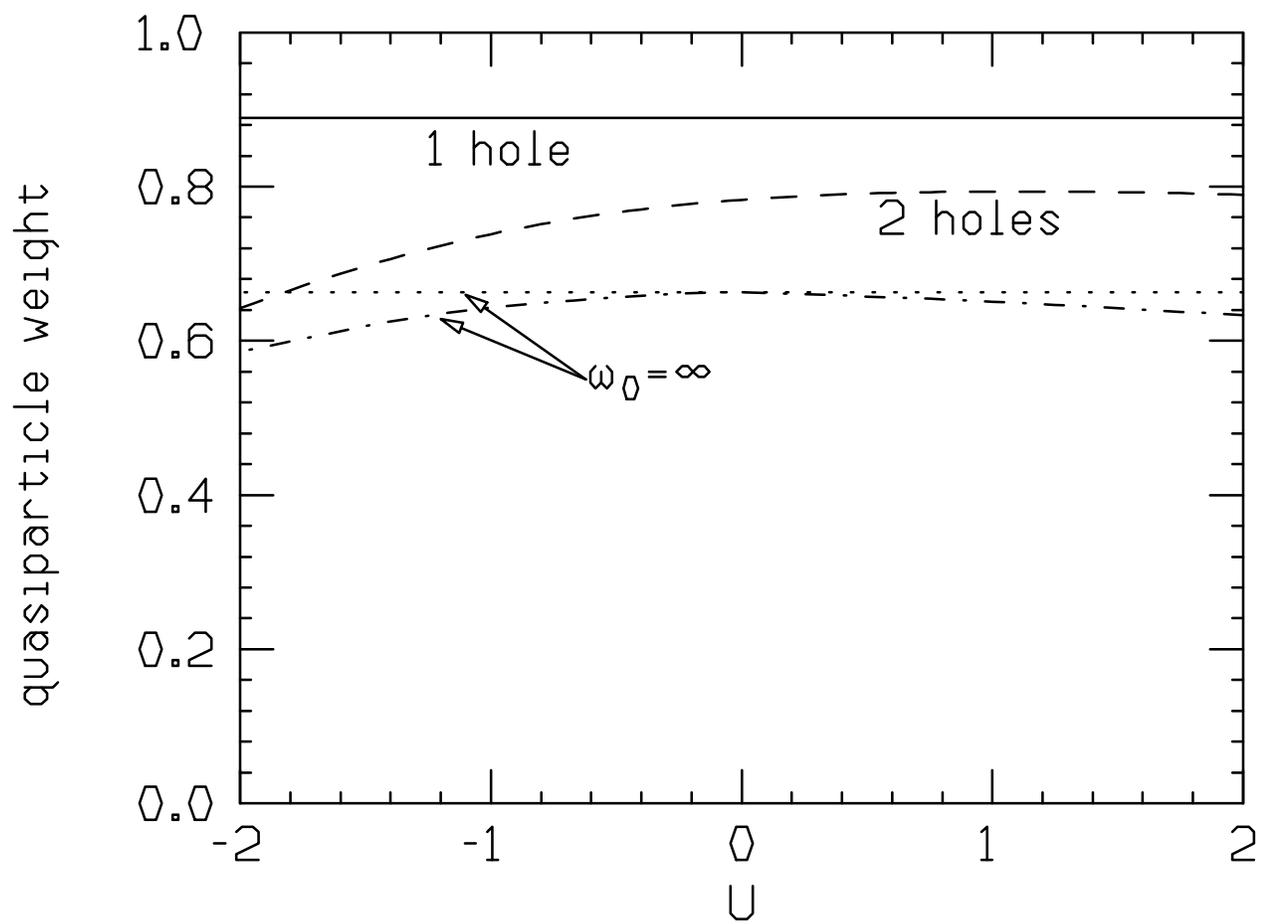
\begin{figure}
\caption {Quasiparticle weights versus $U$ for the electron-hole symmetric
model  with two holes (dashed line)
and with one hole (full line). The corresponding quasiparticle weights
in the antiadiabatic limit are shown as the dash-dotted and dotted lines
respectively.
}
\label{Fig. 13}
\end{figure}
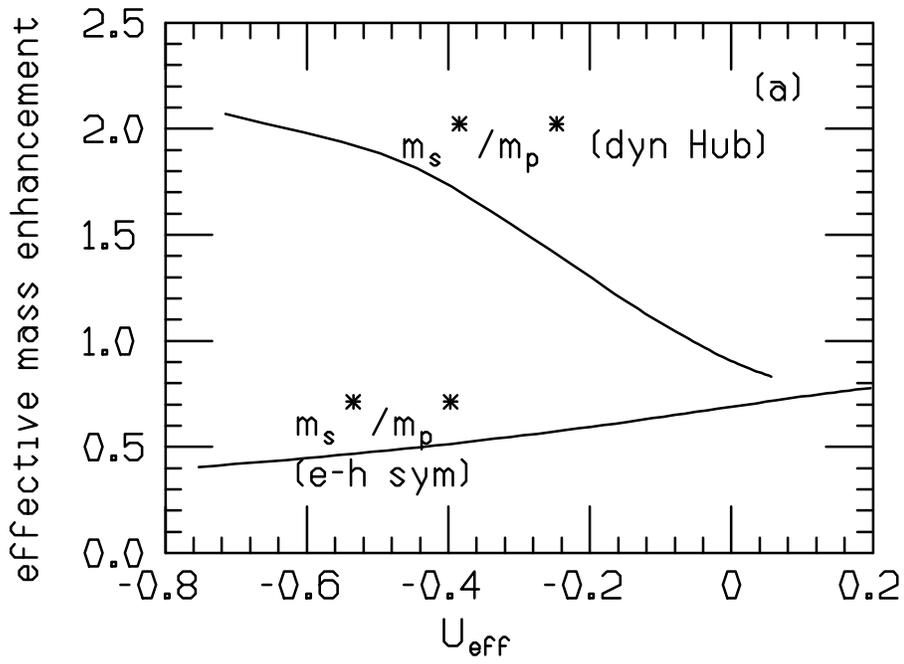
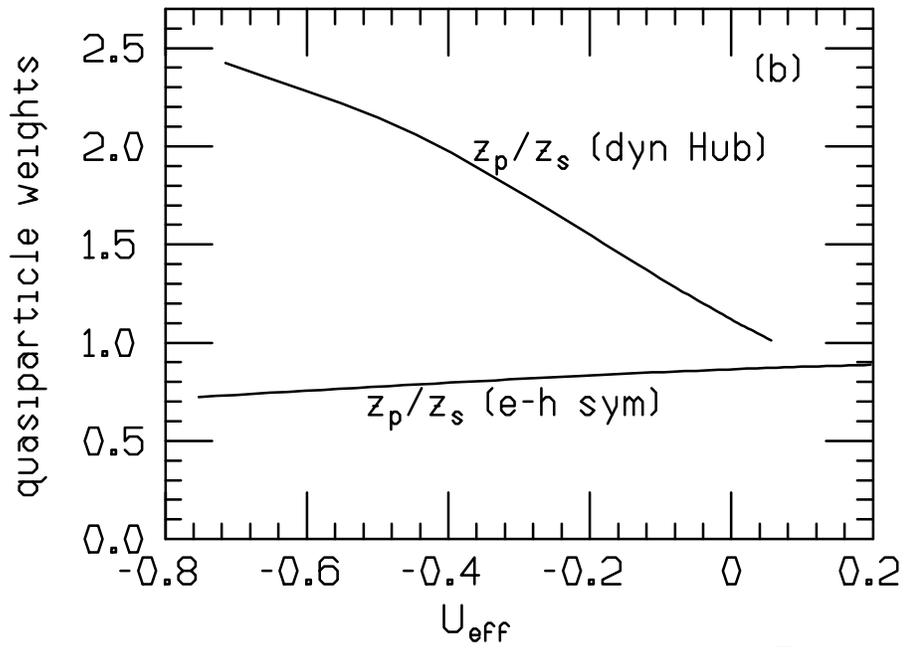
\begin{figure}
\caption {(a) Ratio of effective mass of a single hole to the
effective mass of a hole in a pair, calculated from the ratio of
intra-band kinetic energies, versus effective interaction $U_{eff}$ 
(Eq. (21)) for both models. For the dynamic Hubbard model this ratio
increases as $U_{eff}$ decreases and the pair forms, while for the
electron-hole symmetric model it decreases. (b) Ratio of quasiparticle
weights of a hole in the system with two holes to the quasiparticle weight
of the single hole, versus effective interaction. In the dynamic
Hubbard model this ratio increases as $U_{eff}$ decreases and the
pair forms, in the electron-hole symmmetric model it decreases.
}
\label{Fig. 14}
\end{figure}
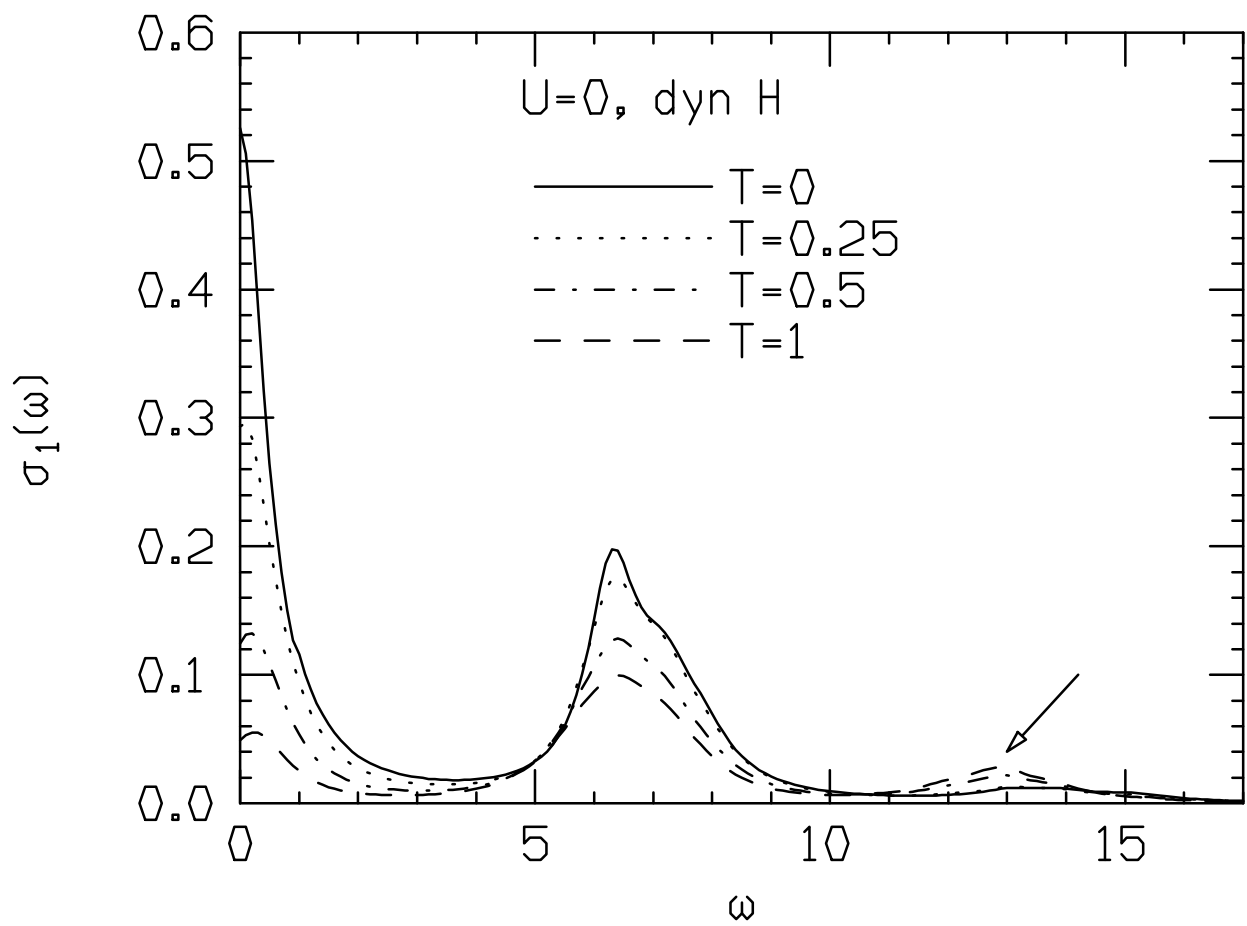
\begin{figure}
\caption {Optical conductivity of the dynamic Hubbard model for
two holes and on-site repulsion $U=0$ for various temperatures
(in units of the bare hopping $t$). Note the transfer
of spectral weight from high frequencies ($\omega > 10$) to low frequencies
as the temperature is lowered.
}
\label{Fig. 15}
\end{figure}

\widetext

\begin{table}
\caption{Optical transitions, possible states of auxiliary spin. The ground state and 
excited state of the auxiliary spin at a site with $n$ holes are denoted
by $|n>$ and $|\bar{n}>$ respectively. The four left  columns correspond to
transitions involving a single hole hopping between neighboring sites,
 $|\uparrow>|0>\rightarrow|0>|\uparrow>$;
the four right columns correspond to transitions involving two holes at neighboring sites, 
$|\uparrow>|\downarrow>\rightarrow |0> |\uparrow\downarrow>$.}
\begin{tabular}{ccccccccc}
initial state&final state& weight & energy &initial state&final state& weight & energy &weight ratio \cr
 $|\uparrow>|0>$&$|0>|\uparrow>$& &  &
$|\uparrow>|\downarrow>$ & $|0>|\uparrow\downarrow >$ &   &  &   \cr
\tableline
$|1>|0>$  &$|0>|1>$  & $S^2$ &intra-band & $|1>|1>$ & $|0>|2>$ & $S$ & intra-band & $1/S>1$ \cr
  &$|\bar{0}>|1>$  & $S$ &$\omega_0\sqrt{1+g^2}$ &  & $|\bar{0}>|2>$ & $1$ &
 $\omega_0\sqrt{1+g^2}$ & $1/S>1$ \cr
 &$|0>|\bar{1}>$  & $S$ &$\omega_0\sqrt{1+g^2}$ &  & $|0>|\bar{2}>$ & $S^2$ &
 $\omega_0\sqrt{1+g^2}$ & $S<1$ \cr
 &$|\bar{0}>|\bar{1}>$  & $1$ &$2\omega_0\sqrt{1+g^2}$ &  & $|\bar{0}>|\bar{2}>$ & $S$ &
 $2\omega_0\sqrt{1+g^2}$ & $S<1$ \cr
\end{tabular} \end{table}

\begin{references}
\bibitem{schrieffer} J.R. Schrieffer, ``Theory of Superconductivity'',
W.A. Benjamin, New York, 1964.
\bibitem{hole1} J.E. Hirsch, Phys.Lett. A {\bf 134}, 451 (1989).
\bibitem{hole2} J.E. Hirsch and F. Marsiglio, Phys. Rev. B {\bf 39}, 11515 (1989).
\bibitem{undressing} J.E. Hirsch, Phys. Rev. B {\bf 62}, 14487 (2000); {\bf 62 }, 14498 (2000).
\bibitem{dynh1} J.E. Hirsch, Phys. Rev. Lett. {\bf 87}, 206402 (2001).
\bibitem{dynh2} J.E. Hirsch, Phys.Rev.B {\bf 65}, 184502 (2002).
\bibitem{ding} H. Ding et al, Phys. Rev. Lett. {\bf 87}, 227001 (2001).
\bibitem{ando} Y. Ando et al, Phys. Rev. Lett. {\bf 87}, 017001 (2001).
\bibitem{johnson} P.D. Johnson et al, Phys. Rev. Lett. {\bf 87}, 177007 (2001).
\bibitem{basov} D.N. Basov et al, Science {\bf 283}, 49 (1999).
\bibitem{santander} A.F. Santander-Syro et al, cond-mat/0111539 (2001).
\bibitem{marel} H. J. A. Molegraaf, C. Presura, D. van der Marel, P. H. Kes, and M. Li
Science {\bf 295}, 2239 (2002).
\bibitem{qmc} J.E. Hirsch, cond-mat/0201005 (2002), to be published in Phys.Rev. B.
\bibitem{maldague} P. Maldague, Phys. Rev. B {\bf 16}, 2437 (1977).
\bibitem{apparent} J.E. Hirsch, Physica C {\bf199}, 305 (1992).
\bibitem{london} J. E. Hirsch and F. Marsiglio, Phys. Rev. B\ {\bf 45}, 4807
(1992).
\bibitem{hanke} J. Wagner, W. Hanke and D.J. Scalapino, Phys.Rev. B {\bf 43}, 10517 (1991).
\bibitem{bose} J.E. Hirsch, Physica C {\bf 179}, 317 (1991).
\bibitem{mahan} G.D. Mahan, ``Many-Particle Physics'' (Plenum, New York, 1981).
\bibitem{jarrell} M. Jarell, Phys.Rev.Lett. {\bf 69}, 168 (1992).
\bibitem{yusof} Z.M. Yusof et al, Phys.Rev.Lett. {\bf 88}, 167006 (2002).
\bibitem{coherence} F. Marsiglio and J.E. Hirsch, Phys.Rev. B {\bf 44}, 11960 (1991)
\bibitem{fugol} I. Fugol et al, Solid St. Comm. {\bf 86}, 385 (1993).
\bibitem{gao} F. Gao, D.B. Romero, D.B. Tanner, J. Talvacchio and 
M.G. Forrester, Phys.Rev. B {\bf 47}, 1036 (1993).
\bibitem{uchida} S.Uchida, T.Ido, H.Takagi, T. Arima, Y. Tokura and
S. Tajima,  Phys. Rev. B {\bf 43}, 7942 (1991).
\bibitem{color} J.E. Hirsch, Physica C {\bf 201}, 347 (1992).
\bibitem{kiv} E.W. Carlson,D.Orgad, S.A.Kivelson and V.J. Emery,
Phys.Rev. B {\bf 62}, 3422 (2000).
\bibitem{and} P.W. Anderson,   Physica C {\bf 341-348}, 9 (2000).
\bibitem{narlikar}J.E. Hirsch,  cond-mat/0106310 (2001), 
to be published in "Studies of High Temperature Superconductors", Vol. 38, ed. by A. Narlikar, 
Nova Sci. Pub., New York.




\end{references}
 \end{document}